\documentclass[twocolumn,iop,apj]{aastex63}
\usepackage{amsfonts,amsmath,graphicx,natbib,url,hyperref,nicefrac}
\usepackage{amssymb, verbatim, subfigure, paralist, soul, comment}
\usepackage{graphicx}
\usepackage{wrapfig}
\usepackage{verbatim}
\usepackage{placeins}
\usepackage{lineno}
\usepackage{tikz}
\usepackage{color}

\usepackage{overpic}
\usepackage{multirow}
\usepackage{float}

\usepackage[T1]{fontenc}
\usepackage{aecompl}
\usepackage{calligra}
\DeclareMathAlphabet{\mathcalligra}{T1}{calligra}{m}{n}
\DeclareFontShape{T1}{calligra}{m}{n}{<->s*[2.2]callig15}{}

\usepackage{ulem}
\usepackage{cancel}
\usepackage{hyperref}
\hypersetup{
    pdfnewwindow=true,      % links in new window
    colorlinks=true,       % false: boxed links; true: colored links
    linkcolor=blue,          % color of internal links
    citecolor=blue,        % color of links to bibliography
    filecolor=blue,      % color of file links
    urlcolor=blue           % color of external links
}

\def\apjl{ApJL}               % Astrophysical Journal, Letters
\def\apjs{ApJS}               % Astrophysical Journal, Supplement
\def\aap{A\&A}                % Astronomy and Astrophysics
\defcitealias{O99}{O99}
\defcitealias{Paper1}{Paper I}

\DeclareMathAlphabet{\mathcalligra}{T1}{calligra}{m}{n}
\DeclareFontShape{T1}{calligra}{m}{n}{<->s*[2.2]callig15}{}

\hypersetup{colorlinks=true, urlcolor=blue, citecolor=blue}

\usepackage{color}

\defcitealias{Bromberg+2011ApJ}{B11}

\newcommand{\OK}[1]{\textcolor{red}{[OK]}}

\usepackage{comment}

\shorttitle{} 
\shortauthors{O'Neill et al.}

\begin{document}

\title[]{Gaseous Dynamical Friction on Hyperbolic Scatterings}

\author[0000-0002-1382-3802]{David O'Neill}
\affiliation{Niels Bohr International Academy, Niels Bohr Institute, Blegdamsvej 17, DK-2100 Copenhagen Ø, Denmark}
\email{david.oneill@nbi.ku.dk}

\author[0000-0002-1271-6247]{Daniel J. D'Orazio}
\affiliation{Space Telescope Science Institute, 3700 San Martin Drive, Baltimore , MD 21}
\affiliation{Department of Physics and Astronomy, Johns Hopkins University,
3400 North Charles Street, Baltimore, Maryland 21218, USA}
\affiliation{Niels Bohr International Academy, Niels Bohr Institute, Blegdamsvej 17, DK-2100 Copenhagen Ø, Denmark}
\email{dorazio@stsci.edu}

\author[0000-0001-8716-3563
]{Martin E. Pessah}
\affiliation{Niels Bohr International Academy, Niels Bohr Institute, Blegdamsvej 17, DK-2100 Copenhagen Ø, Denmark}
\affiliation{School of Natural Sciences, Institute for Advanced Study, 1 Einstein Drive, Princeton, NJ 08540, USA}
\email{mpessah@nbi.ku.dk}

\begin{abstract}
We present a study of equal-mass hyperbolic encounters, embedded in a uniform gaseous medium. Using linear perturbation theory, we calculate the density wakes excited by these perturbers and compute the resulting forces exerted on them by the gas. We compute the changes to orbital energy, orbital angular momentum and apsidal precession across a wide range of eccentrities and pericenter Mach numbers. 
{We identify six distinct classes of hyperbolic orbits, differing through their wake structure and subsequent orbital evolution.} 
We find the gas to always dissipate orbital energy, leading to smaller semi-major axes and higher pericenter Mach numbers. The orbital angular momentum can either increase or decrease, whereas we typically find the orbital eccentricity to be damped, promoting supersonic gas-captures. Additionally, we find that the force exerted by the gas is not strictly frictional -- particularly for asymptotically subsonic trajectories. 
{Therefore, despite the orbit-integrated changes to orbital parameters being similar to those predicted by the \cite{O99} prescription, the time evolution of the density wakes and the instantaneous forces exerted on the perturbers are significantly different. 
}
\\
\end{abstract}

\section{Introduction}
Dynamical friction is a process in which a massive perturber can exchange both energy and momentum with a medium in which it is embedded. The gravitational potential of a moving body excites the medium, leading to the formation of a dense, asymmetric wake. The gravitational force exerted by this wake can alter the dynamical evolution of the perturber, making it an essential process across a wide range of astrophysical systems.\\

Initially introduced for a stellar, collisionless system, \cite{Chandrasekhar_DF} derived the classical dynamical friction force which has subsequently enjoyed applications in galactic astrophysics \citep[eg.][]{ColisionlessGalaxy,1989MNRAS.239..549W, 1983ApJ...264..364L,hashimoto2003circularize}, dense stellar clusters \citep[][]{2004Natur.428..724P, 2010MNRAS.401.2268A}, planetary migration \citep[][]{Del_Popolo_2003} and Kuiper belt dynamics \citep{goldreich2002formation}. By considering a gaseous, collisional environment and a perturber in constant rectilinear motion, \cite{O99} (hereafter \citetalias{O99}) derived an analytical formula for this drag force as a function of the perturber's Mach number $\mathcal{M}$. This drag force is maximal when the perturber moves at the speed of sound ($\mathcal{M}=1$) and scales as $F_\mathrm{GDF}\propto 1/\mathcal{M}^2$ for supersonic trajectories. This form has since become a prominent way to incorporate the feedback of gas on a given trajectory \citep[e.g.][]{Muto11,Planetesimals,GDF_inGlobularClusters,CommonEnvelope,BinaryFormationAGN,SanchezSalcedo2019,GDF_BinaryFormationMechanism,Qian2024,2020MNRAS.499.2608F,Rozner25, Kummer25, Spieksma25}. While the analytical nature of the \citetalias{O99} model enables a wide utility, it relies on the simplified assumption of constant speed, rectilinear motion. Depending on the intricate orbital dynamics of an astrophysical system, it is unclear whether the \citetalias{O99} prescription accurately captures the subsequent orbital evolution for massive perturbers embedded in gaseous environments.\pagebreak

Previous studies have investigated the gaseous dynamical friction (GDF) force for more relevant astrophysical trajectories. \cite{KK07} found that perturbers on circular orbits experience a force with both radial and azimuthal components, where the radial one is subdominant for low Mach numbers ($\mathcal{M}\leq 2.2$) and the azimuthal one is well approximated by the \citetalias{O99} model. More recently, the GDF force experienced by elliptical Keplerian trajectories has been studied by \cite{Paper1} (hereafter \citetalias{Paper1}), \cite{AnalyticalElliptical} and \cite{Sz_lgy_n_2022} finding the interaction with the gaseous medium to lead to the decay of the semi-major axis while increasing the orbital eccentricity in the supersonic regime ($\mathcal{M}>1$) and decreasing the orbital eccentricity in the subsonic regime ($\mathcal{M}<1$). Bound orbits embedded in a homogeneous, static gas are thus expected to evolve into a highly eccentric, supersonic state \citepalias{Paper1}. In contrast, the linear theory describing hyperbolic scatterings in a gaseous medium remains unexplored.\\

Gaseous environments provide a means of energy dissipation for two-body scatterings. Previous studies, combining both analytical and numerical methods have found that the efficiency of binary formation increases in gaseous media \citep{RowanAGN, Rozner2023, DodiciTremaine, Whitehead2024, Qian2024}  although $(i)$ semi-analytical models are reliant on simplified assumptions of gas dynamics and $(ii)$ numerical simulations are more selective across their parameter space. To address these challenges, we present a semi-analytical model that accurately captures the interaction between hyperbolic scattering and an external gaseous medium to linear order, striking a balance between methods $(i)$ and $(ii)$. We investigate the resulting changes in energy, momentum and argument of pericenter across a wide parameter space.\\

This paper is organised as follows. In Section~\ref{sec:Methods} we introduce our analytical and numerical methods, followed by our results in Section~\ref{sec:Results}, where we discuss density wake morphology, force profiles and orbital evolution. Finally, in Section~\ref{sec:Conclusion} we present our conclusions, potential applications and a list of caveats.

\section{Methods}\label{Methods}
\label{sec:Methods}

\subsection{Analytical Approach}
\begin{figure}[t]
    \centering
    \includegraphics[width = 0.48\textwidth]{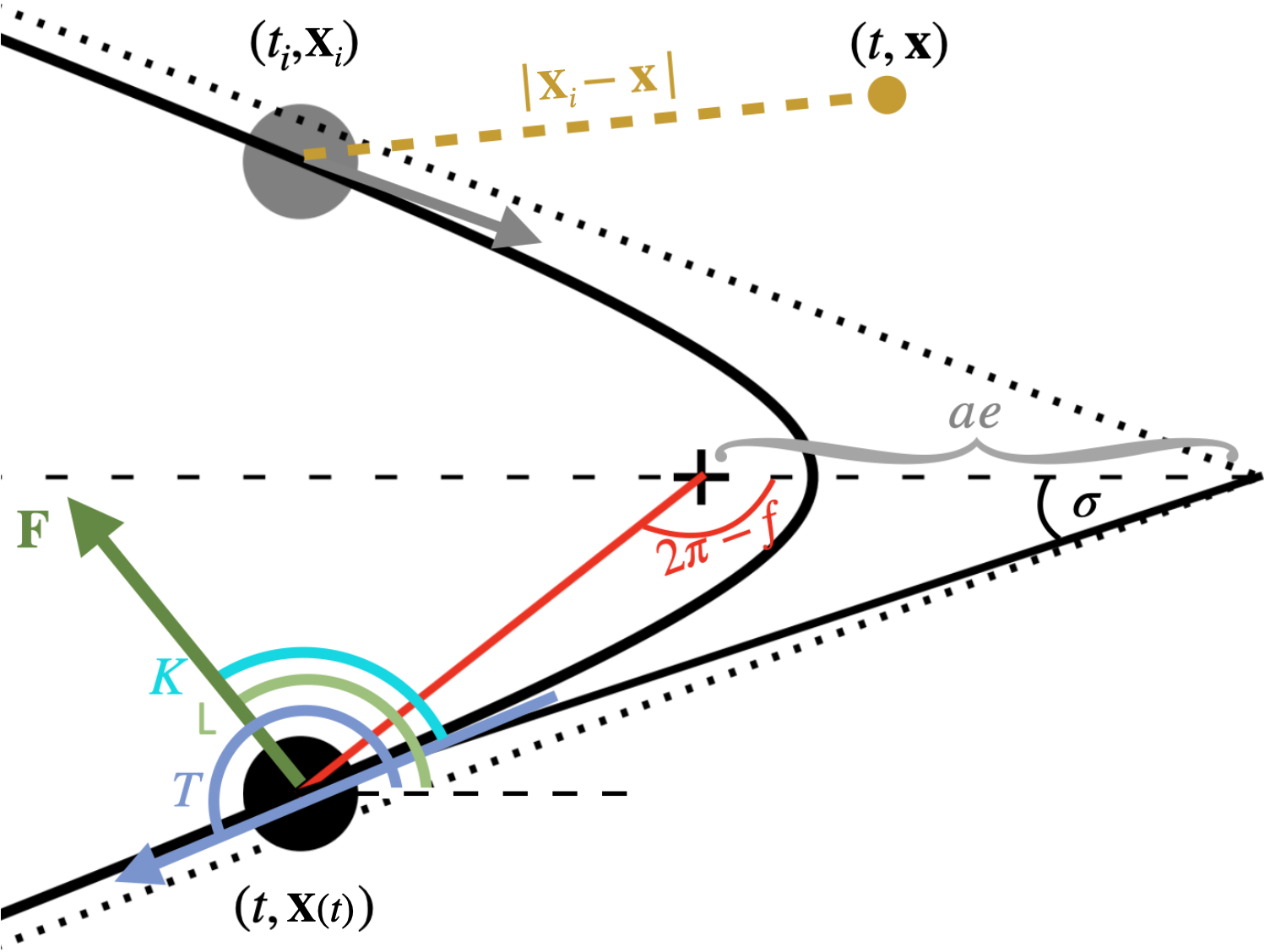}
    \caption{Illustration of the relevant quantities and coordinates considered in this study. The solid black curve is the fixed hyperbolic trajectory followed by the perturber, with it's current location marked by the black circle. The black cross lies at the focus of the hyperbola and the dotted lines are it's asymptotes. The grey circle depicts the perturber's location at some previous time $t_i$, where we can envision a sound wave propagating (yellow) to some point $\mathbf{x}$ in the fluid exciting a density perturbation. The green arrow represents the GDF force, the blue arrow is the tangent line and the coordinate transformation angles $L, K, T, f, \sigma$ are depicted above. 
    }
    \label{fig:Schematic}
\end{figure}

Our approach follows the same general considerations outlined in \citetalias{Paper1}. 
We begin by investigating the fluid's response to the presence of a perturbing body of mass $m$. We assume that the gas is infinite in extent, while initially being both static and homogeneous with density $\rho_0$. The introduction of a massive point particle perturber exerts a gravitational force on the gas, thereby inducing perturbations to density $\alpha$ and velocity $\boldsymbol{\beta}$ of the form, 
\begin{align}
    \rho_\mathrm{gas}(t,\mathbf{x}) &= \rho_0\left[1+\alpha(t,\mathbf{x})\right]\\
    \mathbf{v}_\mathrm{gas}(t,\mathbf{x}) &= c_\mathrm{s}\boldsymbol{\beta}(t,\mathbf{x})
\end{align} 
where $c_\mathrm{s}$ is the sound speed in the medium. Following the same analytical approach as \citetalias{O99}, we assume that both $\alpha,\boldsymbol{\beta}\ll 1$ permitting a linear expansion of the continuity and momentum equations,
\begin{align}
    \frac{1}{c_\mathrm{s}}\frac{\partial\alpha}{\partial t} + \nabla \mathbf\cdot{\boldsymbol{\beta}} &= 0\label{LinearMass}\\
    \frac{1}{c_\mathrm{s}}\frac{\partial\mathbf{\boldsymbol{\beta}}}{\partial t} + \nabla \alpha &= -\frac{1}{c_\mathrm{s}^2}\nabla \Phi_{\mathrm{pert}},
    \label{LinearMomentum}
\end{align}
where $\Phi_{\mathrm{pert}}$ is the gravitational potential of the perturber. So as to model the density wake structure created by the perturber, we decouple Eqs.~(\ref{LinearMass}) and \ref{LinearMomentum}, isolating the $\alpha$ dependence as 
\begin{equation}
    {\partial^2_t\alpha} - c_\mathrm{s}^2\nabla^2\alpha = \nabla^2\Phi_{\mathrm{pert}},
    \label{PerturbedFluid}
\end{equation}
which we identify as a wave equation sourced by the gravitational potential of the perturber. Naturally, the Poisson Equation, $\nabla^2\Phi_\mathrm{pert} = 4\pi G \rho_\mathrm{pert}(t,\mathbf{x})$, with $G$ the gravitational constant, can be used to relate the source in Eq.~(\ref{PerturbedFluid}) to the world-line density of the perturbing body, which we will assume to be prescribed a priori. The solution to Eq.~(\ref{PerturbedFluid}) can thus be expressed as a Green's function \citep{jackson1999classical},
\begin{align}
    \alpha(t,\mathbf{x}) = \frac{4\pi G}{c_\mathrm{s}^2} \int dt' \, d\mathbf{x}' \, \left[\frac{\delta(t' - t + |\mathbf{x}' - \mathbf{x}| / c_\mathrm{s})}{4\pi |\mathbf{x}' - \mathbf{x}|}\right]\nonumber \\ 
    \times \rho_{\mathrm{pert}}(t',\mathbf{x}') H\left[\Omega(t' - t_-)\right],
    \label{GreensFunctionSolution}
\end{align}
where $\Omega = \sqrt{G\mu/a^3}$ is the mean angular motion for a perturber under the gravitational influence of a point mass $\mu$. We envision sound waves being launched from the perturber at coordinates $(t',\mathbf{x}')$, propagating outward into the medium and sourcing density perturbations in the field $\alpha(t,\mathbf{x})$ (see Fig.~\ref{fig:Schematic} for an illustration). Here, $H\left[\Omega(t' - t_-)\right]$ is a heaviside function which ensures that the perturber has only interacted with the medium since some initial time $t_-<0$.\\

For a point-particle perturber with mass $m$, the density world-line can be written as
\begin{equation}
    \rho_\mathrm{pert}(t',\mathbf{x}') = m\delta\left[\mathbf{X}(t') - \mathbf{x}'\right]
    \label{WorldlineDensity}
\end{equation}
where $\mathbf{X}(t)$ is the prescribed trajectory of the perturber. By inserting the world-line density from Eq.~(\ref{WorldlineDensity}) into Eq.~(\ref{GreensFunctionSolution}) we find
\begin{align}
    \alpha(t,\mathbf{x}) = \frac{Gm}{c_\mathrm{s}^2} \int dt' d\mathbf{x}' \left[\frac{\delta\left(t'-t+|\mathbf{x}'-\mathbf{x}|/c_\mathrm{s}\right)}{|\mathbf{x}'-\mathbf{x}|}\right]\\\times \delta\left[\mathbf{X}(t') - \mathbf{x}'\right]H\left[\Omega(t' - t_-)\right].
 \end{align}
 Next, we perform the spatial integral over $\mathbf{x}'$ and express the $t'$ Dirac delta as a sum over roots,
\begin{equation}
    \alpha(t,\mathbf{x}) = \frac{Gm}{c_\mathrm{s}^2} \int dt' \frac{\sum_i\delta(t'-t_i)H\left[\Omega(t' - t_-)\right]}{\left||\mathbf{X}(t')-\mathbf{x}| + \left(\mathbf{X}(t')-\mathbf{x}\right)\cdot\frac{\mathbf{v}(t')}{c_\mathrm{s}}\right|},
\end{equation}
where $\mathbf{v}(t)$ is the velocity of the peturber at time $t$. The density perturbation sourced by a point mass perturber on a fixed trajectory is thus given by
\begin{equation}
    \alpha(t,\mathbf{x}) = \frac{Gm}{c_\mathrm{s}^2}\sum_i \frac{H\left[\Omega(t_i - t_-)\right]}{\left||\mathbf{X}_i-\mathbf{x}| + (\mathbf{X}_i-\mathbf{x})\cdot{\mathbf{v}_i/c_\mathrm{s}}\right|},
    \label{GeneralSolutionAlpha} 
\end{equation}
where we have introduced the shorthand $\mathbf{X}_i \equiv \mathbf{X}(t_i)$ and $\mathbf{v}_i \equiv \mathbf{v}(t_i)$. The roots $i$ satisfy $t_i - t + |\mathbf{X}_i - \mathbf{x}|/c_\mathrm{s}=0$, physically corresponding to previous locations of the perturber where a sound wave was launched so that it will arrive at some point $\mathbf{x}$ in the fluid, at current time $t$.

\subsection{Hyperbolic Trajectories}
\label{sec:HyperbolicTrajectories}
In this paper, we seek specific solutions of Eq.~(\ref{GeneralSolutionAlpha}), describing a perturbing mass on a prescribed hyperbolic, Keplerian trajectory. In doing so, we employ a hyperbolic coordinate system $(a,\sigma,z)$ useful for parameterising the perturber's motion. In this coordinate system, $\sigma(t)$ is the hyperbolic eccentric anomaly and $z$ is the out-of-plane Cartesian coordinate, namely, 
\begin{equation}
    \mathbf{X}(t) = (-a\cosh{\sigma}(t), -a\sqrt{e^2-1}\sinh{\sigma}(t),z),
\label{TrajectoryPrescribed}
\end{equation}
where we adopt the convention of a positive semi-major axis $a>0$, assumed to be fixed. Here, $e>1$ is the eccentricity, which we also assume to be fixed. Furthermore, without loss of generality we can restrict our analysis to trajectories only in the $z = 0$ plane, meaning the only free parameter is the hyperbolic eccentric anomaly $\sigma$ which we can parameterise via Kepler's equation as\begin{equation}
    -\sigma(t) + e\sinh\sigma(t) = \Omega t,
    \label{KeplerEq}
\end{equation}
Although the perturber is assumed to follow a prescribed Keplerian trajectory, we will not assume the existence of a central body of mass $\mu$ responsible for its dynamics. Instead, for the sake of building intuition behind the gas dynamics, we will only consider the wake created by a single perturber, whose trajectory is given by Eq.(\ref{TrajectoryPrescribed}). Moreover, in Section \ref{sec:DoublePerturbers} we will superimpose two single perturber density wakes in modelling an equal mass scattering, where both perturbers orbit the common center of mass of the system.\\

We can write Eq.~(\ref{GeneralSolutionAlpha}) as a product between a dimensional scale times a dimensionless factor $\mathcal{D}_s$, to be interpreted as the dimensionless, single perturber density wake,
\begin{align}
    \mathcal{D}_s\left(\Omega t,{\mathbf{x}}/{a},{\mathbf{X}}/{a},{\mathbf{v}}/{c_\mathrm{s}}\right) &= \sum_i \frac{H\left[\Omega(t_i - t_-)\right]}{\left|\frac{|\mathbf{X}_i-\mathbf{x}|}{a} + \frac{(\mathbf{X}_i-\mathbf{x})}{a}\cdot\frac{\mathbf{v}_i}{c_\mathrm{s}}\right|},\nonumber\\
    \alpha(t,\mathbf{x}) &=\frac{Gm}{ac_\mathrm{s}}\mathcal{D}_s\left(\Omega t,{\mathbf{x}}/{a},{\mathbf{X}}/{a},{\mathbf{v}}/{c_\mathrm{s}}\right).
    \label{DimensionsAndUnits}
\end{align}
In Eq.~(\ref{DimensionsAndUnits}), we have used the semi-major axis $a$ to define the unit of length and we use the mean angular motion $\Omega$ to define an (inverse) unit of time $T=2\pi[\Omega^{-1}]$. In this unit system, the asymptotic Mach number of the perturber is simply given by $\mathcal{M}_\infty = a \Omega / c_\mathrm{s}$, or more generally, the Mach number as a function of the hyperbolic anomaly is given by
\begin{equation}
    \mathcal{M}(\sigma) = \mathcal{M}_\infty \sqrt{\frac{e\cosh\sigma+1}{e\cosh\sigma-1}}.
    \label{HyperbolicVelocity}
\end{equation}
\subsection{Double Perturbers: A Point-Particle Scattering}
\label{sec:DoublePerturbers}
The linear nature of our approach enables solutions to be constructed as the weighted sums of single perturber density wakes. The inclusion of a second massive perturber allows for a consistent treatment of the orbital dynamics (Section \ref{sec:HyperbolicTrajectories}) as both bodies orbit the common center of mass of the system. Accordingly, the density wake $\alpha_\mathrm{e}$ sourced by an encounter of two point masses $m_1$ (primary) and $m_2$ (secondary) is given by the sum,
\begin{align}
    \alpha_\mathrm{e}(t,\mathbf{x}) &= \frac{Gm_1}{a_1c_\mathrm{s}^2}\mathcal{D}_s\left(\Omega t,{\mathbf{x}}/{a_1},{\mathbf{X}}/{a_1},{q\mathbf{v}_2}/{c_\mathrm{s}}\right)\nonumber \\ &+ \frac{Gm_2}{a_2c_\mathrm{s}^2}\mathcal{D}_s\left(\Omega t,{-\mathbf{x}}/{a_2},-{\mathbf{X}}/{a_2},-{\mathbf{v}_2}/{c_\mathrm{s}}\right),
    \label{LinearSum}
\end{align}
where $q=m_2/m_1$ is the mass ratio, $a_1$ and $a_2$ are the respective semi-major axes of the primary and secondary, and the velocity of the primary is $\mathbf{v}_1=-q\mathbf{v}_2$ around their mutual barycenter. In the \emph{relative frame}, the semi-major axis of the encounter is simply given by the sum $a_\mathrm{e}=a_1+a_2$, which can be used to define the unit of length, while $M = m_1+m_2$ defines the unit of mass. Analogous to Eq.~(\ref{DimensionsAndUnits}), the general form for the dimensionless density wake is,
\begin{align}
    \mathcal{D}\left(\Omega t, \frac{\mathbf{x}}{a_\mathrm{e}},\frac{\mathbf{X}}{a_\mathrm{e}},\frac{\mathbf{v}}{c_\mathrm{s}},q\right) 
     &=\frac{1}{1+q}\sum_i \frac{H\left[\Omega(t_i-t_-)\right]}{\left|\frac{|\mathbf{X}_i-\mathbf{x}|}{a_\mathrm{e}} +\frac{(\mathbf{X}_{i} - \mathbf{x})}{a_\mathrm{e}}\cdot\frac{q\mathbf{v}_{i}}{c_\mathrm{s}}\right|} \nonumber\\
     &+\frac{q}{1+q}\sum_j\frac{H\left[\Omega(t_j-t_-)\right]}{\left|\frac{|\mathbf{X}_{j}-\mathbf{x}|}{a_\mathrm{e}} +\frac{(\mathbf{X}_{j} - \mathbf{x})}{a_\mathrm{e}}\cdot\frac{\mathbf{v}_{j}}{c_\mathrm{s}}\right|}.
     \label{DoublePerturberDimensionlessWake}
\end{align}
The density perturbations sourced by an unbound encounter is then given as
\begin{equation}
    \alpha_\mathrm{e}(t,\mathbf{x}) = \frac{GM}{a_\mathrm{e} c_\mathrm{s}^2} \mathcal{D}\left(\Omega t, \frac{\mathbf{x}}{a_\mathrm{e}},\frac{\mathbf{X}}{a_\mathrm{e}},\frac{\mathbf{v}}{c_\mathrm{s}},q\right).
    \label{DoubleDensityWake}
\end{equation}
While Eqs.~(\ref{DoublePerturberDimensionlessWake}) and (\ref{DoubleDensityWake}) are general for any mass ratio, we choose to focus on the case of equal mass scatterings ($q=1$). By this assumption, the forces acting on each perturber are equal (at all times) and, consequently, there is no center of mass drift for $q=1$ scatterings.

\subsection{Dynamical Forces}
\label{sec:FrictionForce}
The dimensionless density perturbation $\alpha_\mathrm{e}(t,\mathbf{x})$ given in Eq.~(\ref{DoubleDensityWake}) will exert a gravitational force on both perturbers. Due to the symmetric nature of an equal mass encounter, we can compute the time-dependent force $\mathbf{F}_\mathrm{G}(t)$ acting on one of the bodies (with current position $\mathbf{x}_\mathrm{p}$) by integrating over the wake,
\begin{align}
    \mathbf{F}_\mathrm{G}(t) &= \frac{GM\rho_0}{2}\int d\mathbf{x}^3 \frac{\alpha_\mathrm{e}(t,\mathbf{x})(\mathbf{x}-\mathbf{X}(t))}{|\mathbf{x}-\mathbf{X}(t)|^3} \label{IntegrationStep} \\
    &= 2\pi \left(\frac{G^2M^2\rho_0}{c_\mathrm{s}^2}\right)\mathcal{F}\left(\Omega t,\mathbf{x}/a_\mathrm{e}, e, \mathcal{M}_p\right),
\label{IntegrationForce}
\end{align}
where $\mathcal{F}\left(\Omega t,\mathbf{x}/a_\mathrm{e}, e, \mathcal{M}_p\right)$ is the dimensionless force computed in the unit system described in Section~\ref{sec:DoublePerturbers}. Following \citetalias{Paper1}, we define coordinate transformations allowing us to express this force in the polar basis given either (a) the force in the Cartesian basis or (b) the force in the tangential basis (see Fig.~\ref{fig:Schematic} for an illustration)
\begin{align}
    \mathrm{(a)}~\mathbf{F}_{\mathrm{G}} &= F\cos(L-f)\hat{r}+F\sin(L-f)\hat{\theta}\label{CartesianPolar}\\
    \mathrm{(b)}~\mathbf{F}_{\mathrm{G}} &= F\cos(T-f-K)\hat{r}-F\sin(T-f-K)\hat{\theta},
    \label{TangentialPolar}
\end{align}
where $F$ is the magnitude of the force, $f$ is the true anomaly, $K$ is the angle of the force relative to the tangent line $T$ and $L = \arctan(F_{y}/F_{x})$ (where $F_{x}$ and $F_{y}$ are the Catesian components of the force vector in Fig.~\ref{fig:Schematic}).\\

The total force experienced on the perturbers is twice that of Eq.~(\ref{IntegrationForce}), and therefore the specific power $\mathcal{P}$, the specific torque $\mathcal{T}$, and the rate of precession $\dot{\omega}$ \citep{Orbital_Elements} induced by the gas on each perturber is given by
\begin{align}
    \mathcal{P} &= \dot{\mathbf{x}}_\mathrm{P}\cdot \frac{2\mathbf{F}_\mathrm{G}}{M} = \left(\frac{4\pi G^2M\rho_0}{c_\mathrm{s}^2}\right) \frac{a_\mathrm{e}\Omega} {2}\mathcal{P}_0\label{Power}\\
    \mathcal{T} &= \mathbf{x}_\mathrm{P}\times \frac{2\mathbf{F}_\mathrm{G}}{M} = \left(\frac{4\pi G^2M\rho_0}{c_\mathrm{s}^2}\right)\frac{a_\mathrm{e}}{2}\mathcal{T}_0,\label{Torque}\\
    \dot{\omega} &= \frac{2}{M} \sqrt{\frac{a_\mathrm{e}(e^2-1)}{GMe^2}}\biggl[\mathbf{F}_{\mathrm{G}}\cdot\hat{r}\cos{f} +\mathbf{F}_{\mathrm{G}}\cdot\hat\phi \sin{f} \nonumber\\ 
    &\times\left(\frac{2+e\cos{f}}{1+e\cos{f}}\right)
    \biggr] = \left(\frac{4\pi G^2M\rho_0}{c_\mathrm{s}^2}\right) \left(\frac{a_\mathrm{e}\Omega}{2}\right)^{-1} \dot{\omega}_0,\label{Precession}
\end{align}
where $\mathcal{P}_0, \mathcal{T}_0$, and $\dot{\omega}_0$ are dimensionless quantities expressed in the unit system described in Section~\ref{sec:DoublePerturbers}. Integrating Eqs.~(\ref{Power}), (\ref{Torque}) and (\ref{Precession}) over a time domain\footnote{We provide more precise definitions of $t_-$ and $t_+$ in Section~\ref{sec:NumericalConsiderations}} $t_-\leq t \leq t_+$
will provide the changes to orbital energy $\Delta E$, angular momentum $\Delta L$, and the gas-induced precession angle $\Delta \omega$,
\begin{equation}
        (\Delta E, \Delta L, \Delta \omega) = M \int_{t_-}^{t_+} (\mathcal{P} , \mathcal{T}, \dot{\omega})~ dt.
        \label{IntegratedOrbitalElements}
\end{equation}
By appropriately scaling the gas density and perturber mass, we can ensure that the change in orbital parameters is small enough such that the assumption of a fixed orbit remains valid. To quantify this scale, we introduce two dimensionless parameters,
\begin{align}
    \mathcal{A} &= \frac{GM}{a_\mathrm{e} c_\mathrm{s}^2},\\
    \tau &= \frac{G\rho_0}{\Omega^2},
\end{align}
where $\mathcal{A}$ is to be interpreted as the non-linearity parameter and $\tau$ as the ratio of orbital to gas free-fall timescales (see \citetalias{Paper1}, section 2.5). We express Eqs.~(\ref{Power}), (\ref{Torque}), and (\ref{Precession}) in terms of these dimensionless parameters,
\begin{align}
    \mathcal{P} &= 2\pi \mathcal{A}\tau \times \Omega^3a_\mathrm{e}^2 \mathcal{P}_0\label{DimlessPower}\\
    \mathcal{T} &= 2\pi \mathcal{A}\tau \times \Omega^2a_\mathrm{e}^2\mathcal{T}_0\label{DimlessTorque} \\
    \dot{\omega} &= 2\pi\mathcal{A}\tau \times \Omega\dot{\omega}_0.
    \label{DimlessOmega}
\end{align}
We will thus assume that $\mathcal{A}\tau \ll 1$, so that the assumption of fixed orbits remains valid -- as the feedback on the trajectory is negligible. However, it is not always possible to enforce \emph{both} $\mathcal{A} , \tau \ll 1$ as the non-linearity parameter scales with the asmyptotic Mach number $\mathcal{A} = 4\mathcal{M}^2_\infty$. Therefore, we note that asymptotically supersonic encounters are not expected to be in the linear regime.

\subsection{The Classification of Trajectories}
\label{sec:Classifications}
We introduce a series of orbital classifications to quantify distinct regions of parameter space through their wake structure and subsequent orbital evolution. Following \citetalias{Paper1}, a trajectory is said to be strictly subsonic if $\mathcal{M}_p<1$, strictly supersonic if $\mathcal{M}_\infty >1$ and otherwise transonic.\\

A novel \emph{subclassification} presents itself for strictly supersonic orbiters. We can distinguish between the orbits which, when scattered, will re-enter their wake and those which will not. Depending on the angle of the incoming Mach cone and the angle of deflection, a simple analytical argument can be made to separate these two classes of orbits. From Eq.~(\ref{TrajectoryPrescribed}) the perturber enters from an asymptotic angle $A$, such that \begin{equation}
    \tan{A} = -\sqrt{e^2-1},
    \label{EntranceAngle}
\end{equation}
which, during the asymptotic approach of pericenter has Mach number $\mathcal{M}_{\infty}$ (from Eq.~\ref{HyperbolicVelocity}). Next, we use this incoming Mach number to compute the half-opening angle, $C$, of the rectilinear Mach cone (see Section 2 \citetalias{O99}),
\begin{equation}
    \sin{C} = \frac{1}{\mathcal{M}_{\infty}}.
    \label{SinC}
\end{equation}
\begin{figure}[t]
    \centering
    \includegraphics[width=\linewidth]{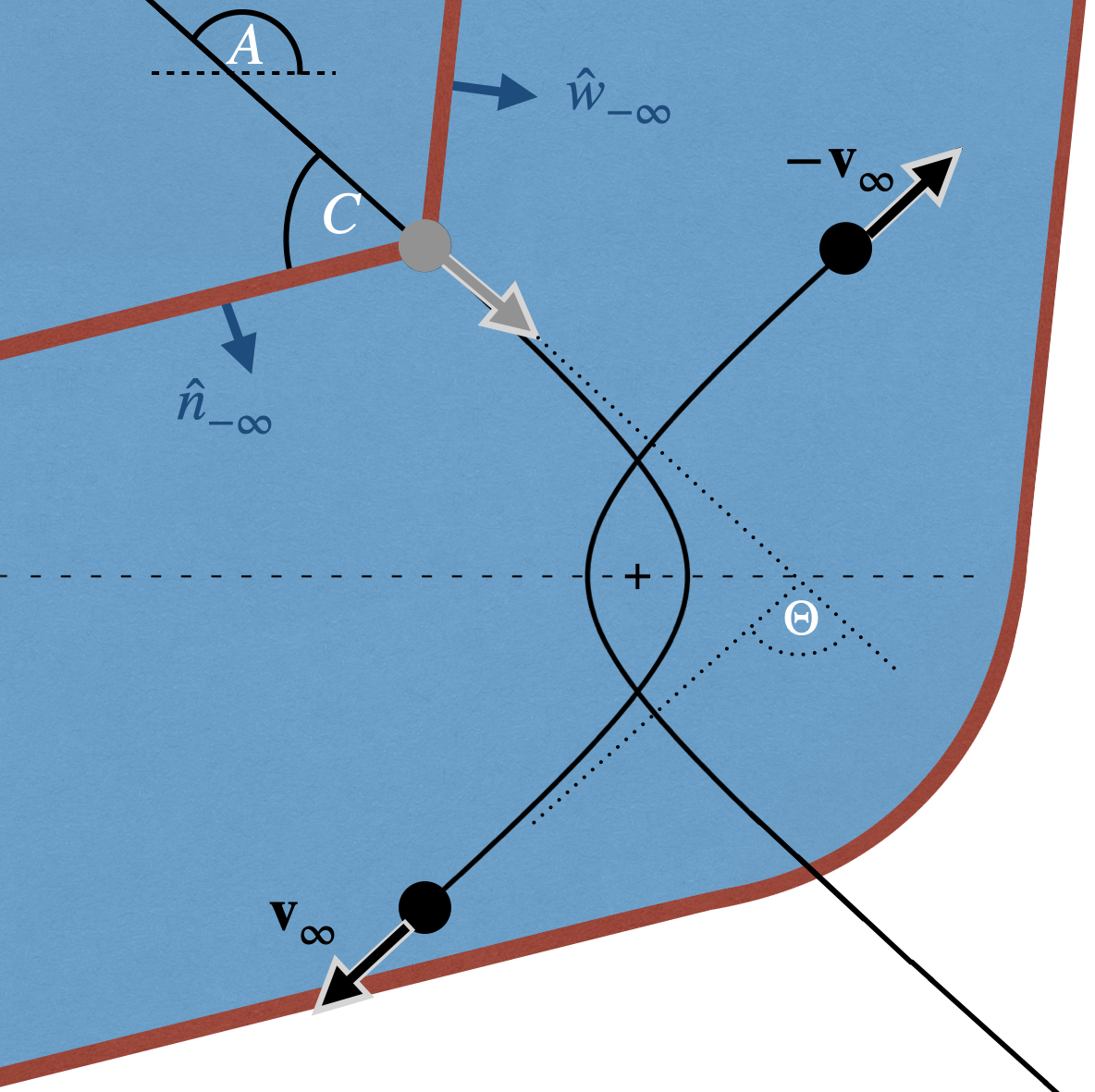}
    \caption{Depending on the orbital eccentricity and pericenter Mach number, both, one or none of the perturbers may escape the wake created during the incoming approach. \emph{Black Lines:} The prescribed trajectories of the massive perturbers. \emph{Blue Shade:} The density wake created by one of the incoming perturbers. \emph{Brown Lines:} The front of the density wake, propagating at the speed of sound $c_\mathrm{s}$.}
    \label{fig:Classifications}
\end{figure}
While the gas dynamics is linear in nature, the Mach cone represents a contact discontinuity in the fluid density, pressure, and velocity which propagates at the speed of sound in the direction $\hat{n}_{-\infty}$ normal to its surface, with
\begin{equation}
    \hat{n}_{-\infty} = -\sin(A+C)\hat{x} +\cos{(A+C)}\hat{y}.
    \label{NormalVector}
\end{equation}
To quantify whether or not an orbit will escape this surface, we define the projected-Mach number $\mathcal{M}_\mathrm{proj}$ as the inner product between the perturber's outgoing velocity (at $t=+\infty$) and the unit vector normal to the incoming Mach cone (in Eq.~\ref{NormalVector}) divided by the sound speed $c_\mathrm{s}$,  
\begin{align}
\mathcal{M}_\mathrm{proj}&\equiv\frac{\mathbf{v}_\infty\cdot{\hat{n}_{-\infty}}}{c_\mathrm{s}} \nonumber \\
&= \frac{1}{e^2}\left[-2 + 2\sqrt{e^2-1}\sqrt{\mathcal{M}_{\infty}^2-1} + e^2\right].
\end{align}
The value of $\mathcal{M}_\mathrm{proj}$ establishes whether or not the perturber escapes the advance of the contact discontinuity. In particular, trajectories with $\mathcal{M}_\mathrm{proj} < 1$ are considered ``embedded'' in nature, whereas the ones with $\mathcal{M}_\mathrm{proj} \geq 1$ are considered ``extracted'' as they will escape to a region of unperturbed gas. The separatrix between these two regimes occurs at some critical pericenter Mach number $\mathcal{M}^\triangleright_{p}$, such that $\mathcal{M}_\mathrm{proj}=1$ for which,
\begin{equation}
    \mathcal{M}^\triangleright_{p} = \frac{e}{e-1}.
    \label{CriticalPericenter}
\end{equation}
We note that in the limit of rectilinear motion $e\to\infty$, this separatrix in Mach number approaches unity $\mathcal{M}_p^\triangleright\to1$. At this ``sonic'' Mach number, a rectilinear motion perturber produces divergent density perturbations $\alpha\gg 1$, as observed by \citetalias{O99}.\\ 

Thus far, we have ignored the Mach cone created by the companion perturber. Following a similar approach, we define the unit vector normal to the Mach cone created by the companion wake $\hat{w}_{-\infty}$
\begin{equation}
    \hat{w}_{-\infty} = - \sin{(A-C)}\hat{x} + \cos{(A-C)}\hat{y},
\end{equation}
which when projected along the asymptotic velocity $\mathbf{v}_{\infty}$ provides a measure of the projected Mach number relative to the companion wake. We find that
\begin{figure}[t]
    \centering
    \includegraphics[width=\linewidth]{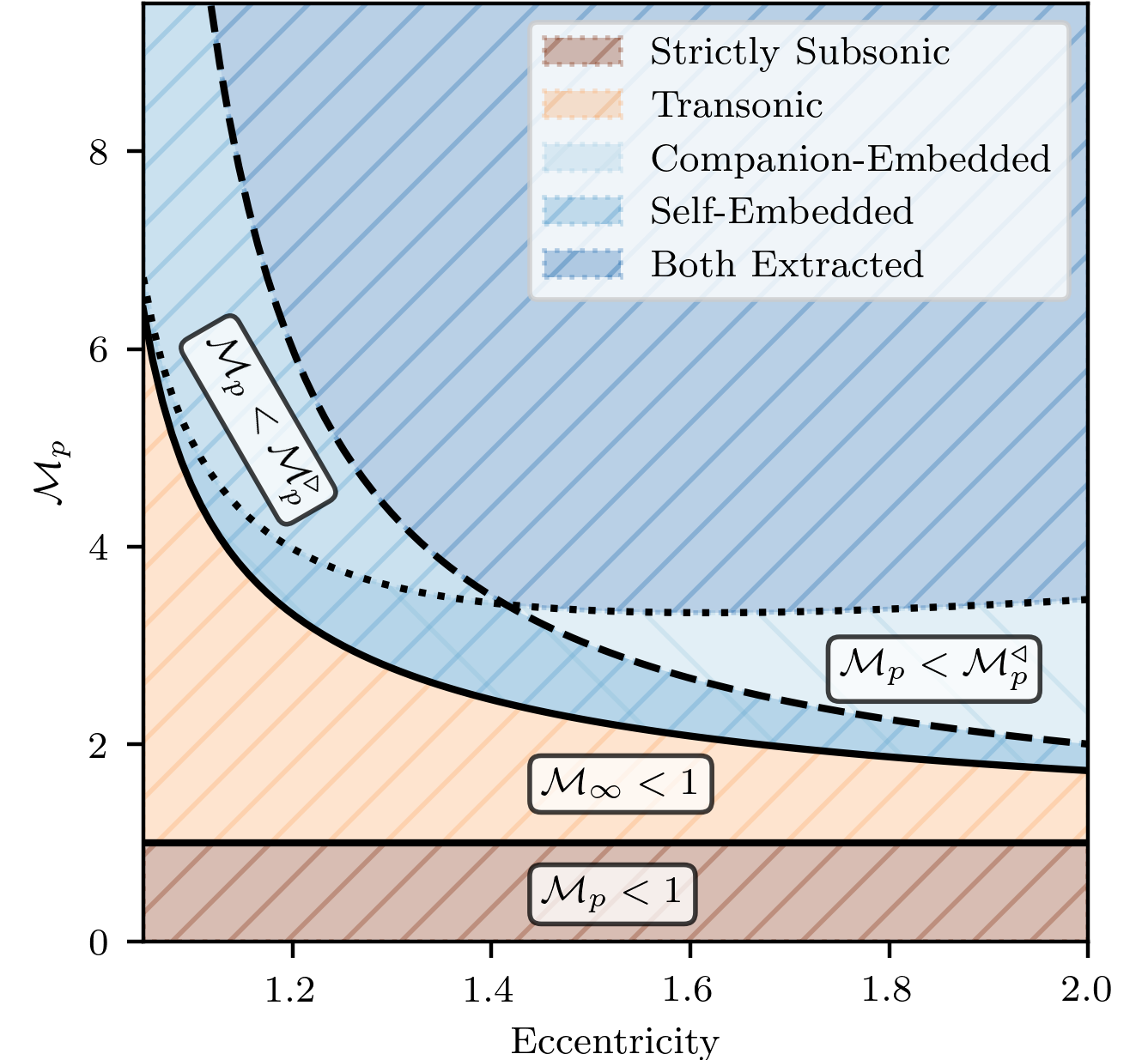}
    \caption{A parameter space spanned by orbital eccentricity and pericenter Mach number. The shaded regions indicate the regions of parameter space according to the orbital classifications introduced in Section~\ref{sec:Classifications}.}
    \label{fig:PartsOfParameterSpace}
\end{figure}

\begin{equation}
    \mathcal{M}^\triangleleft_{p} = e\sqrt{\frac{e+1}{e-1}}, 
\end{equation}
or equivalently, when $\mathcal{M}_{\infty} = e$.\\

In summary, the previous considerations lead to a set of orbital classifications:
\begin{itemize}
    \item \textbf{Strictly Subsonic }$\mathcal{M}_p < 1$ : Both perturbers are always subsonic during the encounter.
    \item \textbf{Transonic }$\mathcal{M}_p \geq 1 > \mathcal{M}_\infty$ : Each perturber will cross the sound barrier twice -- once before and once after pericenter.
    \item \textbf{Self-Embedded }$\mathcal{M}_p < \mathcal{M}^\triangleright_{p}$ : Each perturber is always supersonic but will scatter back into their own wake.
    \item \textbf{Self-Extracted }$\mathcal{M}_p \geq\mathcal{M}^\triangleright_{p}$ : Each perturber is always supersonic and capable of escaping the wake it created during its pericenter approach.
    \item \textbf{Companion-Embedded} $\mathcal{M}_p < \mathcal{M}^\triangleleft_{p}$ : Each perturber is always supersonic although will enter the wake made by its companion.
    \item \textbf{Companion-Extracted} $\mathcal{M}_p \geq \mathcal{M}^\triangleleft_{p}$ : Each perturber is always supersonic and capable of escaping the wake created by its companion.
\end{itemize}

We note that self-extracted and companion-extracted trajectories are mutually independent of one another; a trajectory may belong to one classification without necessarily belonging to the other. Trajectories which belong to both will asymptotically reach unperturbed gas, essentially escaping the ``memory'' of the encounter. Otherwise, the orbit will asymptotically remain embedded inside of a pre-existing wake, continuing to interact with it indefinitely. This subclassification cannot apply to trajectories which are asymptotically subsonic, as the angle $C$ is undefined when $\mathcal{M}_\infty<1$ (see Eq.~\ref{SinC}).

\subsection{Rectilinear Proxies Based on \citetalias{O99}}
\label{sec:RectlinearProxy}
To facilitate direct comparisons with the results obtained using the analytical model derived in \citetalias{O99}, we introduce a ``rectilinear proxy'' density wake.  This proxy wake enables the calculation of a ``proxy force'' exerted on a perturber moving along a hyperbolic path by considering its motion as a sequence of independent, straight-line segments in which the perturber's motion is taken to be constant and rectilinear. Under this approximation for the instantaneous motion, we can obtain the proxy wake structure, as derived in \citetalias{O99}, and compute the gravitational force this wake exerts on the perturber generating it. We refer to this as the ``one-wake proxy force''. Note that, in addition to this force, each perturber undergoing a hyperbolic encounter is also in principle subject to the gravitational force exerted by wake produced by the other perturber. Using the ``rectilinear proxy'' density wakes for each perturber, one can build a ``two-wake proxy force'' taking into account the effects of the wakes generated by both perturbers acting on each of them.\\

It is important to recognize that proxy wakes can differ significantly from those produced by a perturber following a true hyperbolic trajectory. This is because the proxy approach approximates continuous hyperbolic motion as a series of discrete, straight-line steps, each creating an independent wake that instantaneously trails the perturber. As a result, these proxies do not account for the time-dependent evolution of the wake structure and its history, which naturally occurs during a hyperbolic encounter. Consequently, the dynamical evolution of a perturber calculated using proxy wakes (whether using the "one-wake" or "two-wake" method) may deviate from that obtained when the full time history of the wakes is consistently considered along the true hyperbolic path.\\

By construction, the ``one-wake proxy force'' exerted by the rectilinear proxy wake on the perturber producing it is always tangential to its motion. To compare the dynamical evolution of a perturber subject to this ``one-wake proxy force'', we decompose it into radial and azimuthal components using Eq.~(\ref{TangentialPolar}). In order to obtain a continuous evolution, we evaluate the \citetalias{O99} force prescription with a linear interpolation between $\mathcal{M} \in \{0.97, 1.02\}$ (to circumvent the divergent behaviour at $\mathcal{M} = 1$) while varying the Mach number according to Eqs.~(\ref{KeplerEq}) and (\ref{HyperbolicVelocity}). These considerations allow us to compare the results obtained with the proxies built with the widely adopted formulas derived in \citetalias{O99} with the results we obtain for a perturber considering the time history of the wakes they produce.\\

Typical applications of the \citetalias{O99} prescription for perturbing bodies usually include only the effects of the wake generated by the perturber acting on it, neglecting the wake of the companion (i.e., what we referred to as a ``one-wake'' proxy force above). Prominent examples in the literature include: \citet{2015MNRAS.451.2174T, 2016ApJ...820..106G, 2016MNRAS.462.3812T, 2020MNRAS.493.4861G, 2020ApJ...896..113L, 2020ApJ...898...25T, 2021PhRvD.103b3015C, 2022MNRAS.510..531C,2022ApJ...931..149R,Rozner2023,GDF_BinaryFormationMechanism,Qian2024, 2024MNRAS.528.4958W,DodiciTremaine, Kummer25}. In Section~\ref{sec:Results}, we thus consider the ``one-wake'' proxy force as the appropriate benchmark against which we compare our results that naturally account for the time history effects of both wakes acting on each perturber as they undergo the hyperbolic encounter. A more sophisticated treatment of the rectilinear proxy can be achieved by including the companion wake. In Appendix~\ref{sec:AdvancedProxy}, we show the results obtained with this ``two-wake proxy force''. However, we note that this approach does not produce better agreement with our results.\\

In what follows, unless otherwise noted, we will refer to the previously defined ``one-wake'' proxy as simply the ``rectilinear proxy.''

\begin{figure*}[htbp]
    \centering
    \includegraphics[width = \linewidth]{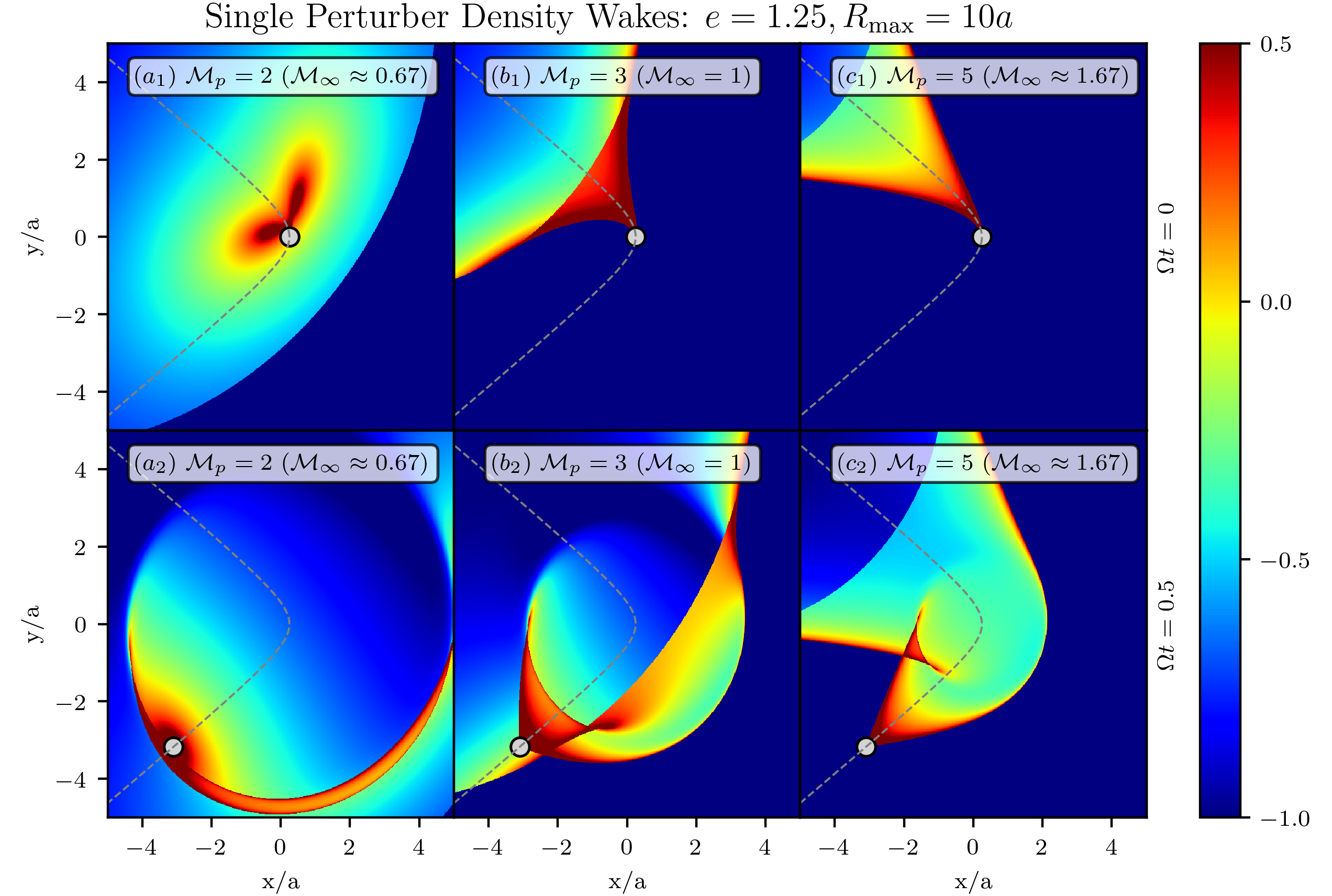}
    \caption{The logarithm of the dimensionless density perturbation $\mathcal{D}_s(\Omega t, \mathbf{x}/a,e,\mathcal{M}_p)$ (Eq.~\ref{DimensionsAndUnits}) for three different Mach numbers, $\mathcal{M}_p \in [2,3,5]$ all with eccentricity $e = 1.25$. The pericenter passage ($t = 0$) is shown on the top row, while the bottom row depicts a later time, $t = 0.5~[2\pi\Omega^{-1}]$. Column $(a)$ is transonic, $(b)$ is self-embedded and $(c)$ is self-extracted (see Section~\ref{sec:Classifications}). Each panel is plotted over a Cartesian domain of size $10a\times 10a$ with grid resolution of $\delta = 0.0167 a$, a medium radius $R_\mathrm{max}=10a$ and thus an initial time $t_- \approx -1.28~[2\pi\Omega^{-1}]$.}
    \label{fig:DensityWakes}
\end{figure*}

\subsection{Numerical Considerations}
\label{sec:NumericalConsiderations}
Characterising the change of orbital parameters over the course of a scattering depends on the length of time over which the perturber interacts with the gas, or equivalently, the size of the medium. To quantify this scale, we define $R_\mathrm{max}$ as the radius of the medium, measured from the barycenter. Thus, given a prescribed trajectory $\mathbf{X}(t)$, we calculate $t_-$ as the time at which the perturber enters the medium (and begins to interact with the gas) and $t_+=|t_-|$ as the time when the perturber exits the medium (ending the interaction with the gas)\footnote{We note, however, that the wakes generated by the perturbers
may extend beyond this scale.}. This allows us to treat $R_\mathrm{max}$ as a free parameter and consider the effects of different sized media.\\

We construct the density wake in Eq.~(\ref{DoublePerturberDimensionlessWake}) iteratively over concentric spherical shells centered at the perturbers location. The surface of each shell contains $2.25\times 10^4$ points, with each step increasing the radius of the shell logarithmically. This iterative process begins at a minimum radius of $r_\mathrm{min} = 0.05 a_e$ and ends once the density perturbations vanish across a shell whose radius exceeds the separation of the two bodies. For a medium size of $R_\mathrm{max} = 100a_e$, this typically consists of $\sim 400$ radial shell iterations before the density wake has finished being constructed.\\

We integrate the density wake (as described in Eq.~\ref{IntegrationStep}) using Simpson's rule and subsequently perform a coordinate transformation (Eq.~\ref{CartesianPolar}) to obtain the dimensionless force in the polar basis. The instantaneous power, torque and the rate of precession exerted by the gas follow from Eqs.~(\ref{Power}),~(\ref{Torque}) and (\ref{Precession}). We repeat this process across a time domain consisting of approximately 30 timesteps within the range $[t_-, t_+]$. The timeseries cadence is increased near pericenter, to ensure that all intricate dynamics are captured comprehensively, providing an accurate model of the dynamical force profile. Integrating the power, torque, and rate of precession provides the changes to the orbital elements of the scattered bodies (Eq.~\ref{IntegratedOrbitalElements}) for a given initial Mach number and orbital eccentricity.\\

The Keplerian trajectory is always prescribed along a hyperbola of fixed eccentricity for mean angular motion of $\Omega = 2\pi[T^{-1}]$ where $T$ is the unit of time. We vary the sound speed $c_\mathrm{s}$ as a means to change the pericenter Mach number and $(t_-,t_+)$ to change the effective size of the medium. To this end, we compute the change in energy, momentum, and the angle of precession for each scattering, across a parameter space spanned by eccentricity $e$, pericenter Mach number $\mathcal{M}_p$, and medium radius $R_\mathrm{max}$.

\section{Results}\label{sec:Results}
\subsection{Gas Wake Morphology}
\label{sec:GasWakeMorphology}
In preparation for two body encounters, we begin by analysing the single perturber density wakes. This ``stepping stone'' is beneficial in building intuition behind the gas dynamics prior to analysing orbital evolution results.\\

In Fig.~\ref{fig:DensityWakes} we present the density wake morphology for trajectories of fixed eccentricity $e=1.25$, while varying the pericenter Mach number $\mathcal{M}_p = \{2, 3, 5\}$. Column $(a)$ illustrates a transonic trajectory, which crosses the sound barrier twice during the encounter (before and after pericenter). Accompanying these sound barrier crossings is the emission of a dense spherical shell -- understood to be the locations at which sound waves accumulate, analogous to the Doppler effect of a source in motion. Both columns $(b)$ and $(c)$ are strictly supersonic, although classified differently according to the self-embedding criterion introduced in Section~\ref{sec:Classifications}. For an eccentricity of $e=1.25$, the separatrix occurs at $\mathcal{M}^\triangleright_{p} = e/(e-1) = 5$, for which the trajectory in column $(b)$ is self-embedded while $(c)$ is self-extracted. This is in agreement with the density wake morphology illustrated in Fig.~\ref{fig:DensityWakes}, where panel $(b_2)$ re-enters the wake while $(c_2)$ does not.\\

\begin{figure}[H]
    \centering
    \includegraphics[width=0.9\linewidth]{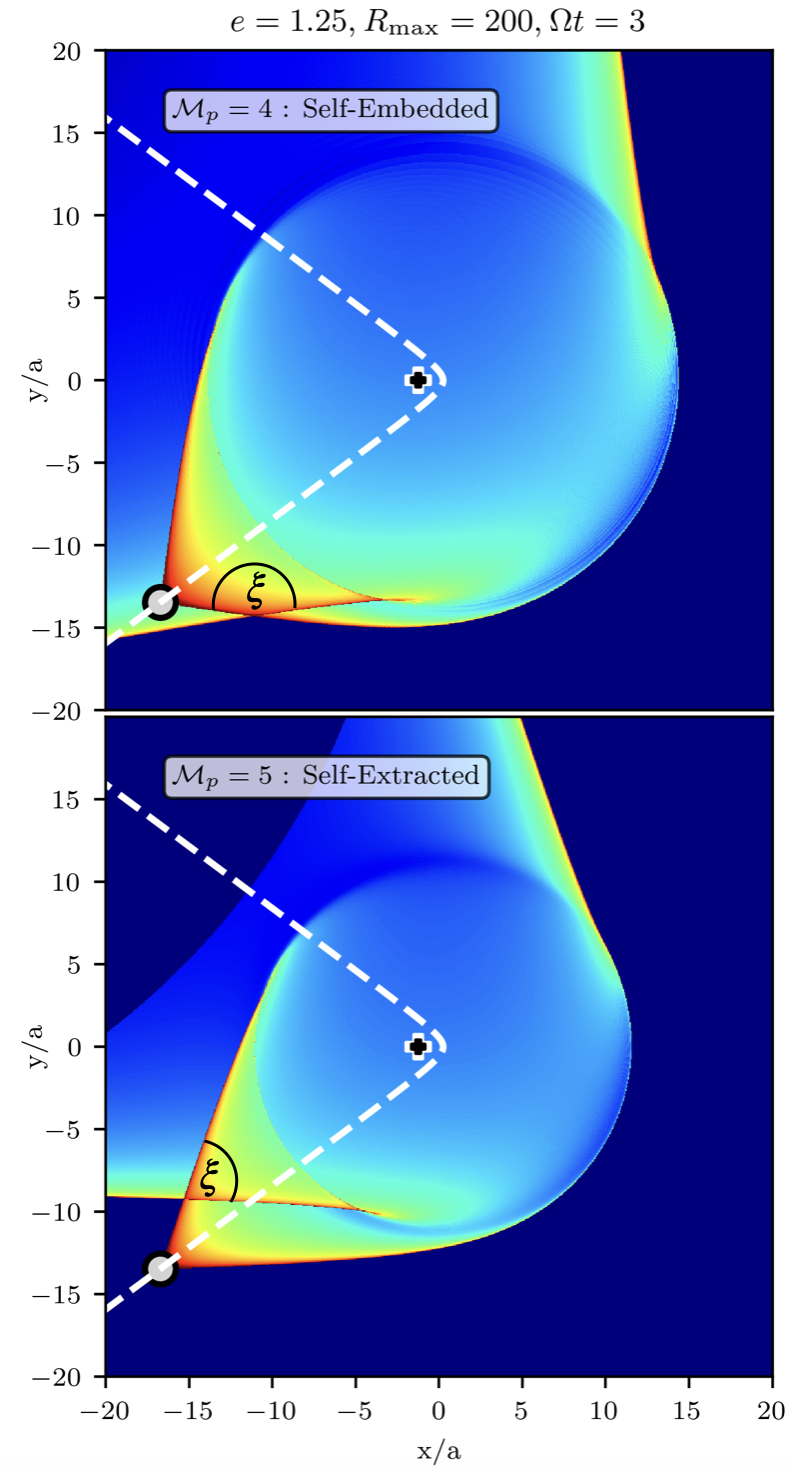}
    \caption{The intersection of the incoming and outgoing Mach cones for $e=1.25$ at time $\Omega t = 3$ over a spatial domain of size $L = 40 \times 40$. Both panels are strictly supersonic, although the top panel has a lower Mach number and is self-embedded whereas the bottom panel has a larger Mach number and is self-extracted.}
    \label{fig:Pinch}
\end{figure}

Spherical shells in the density wake structure are present in all panels of Fig.~\ref{fig:DensityWakes}. There are two distinct mechanisms capable of forming these features. The first mechanism is related to the finite duration of the interaction between the massive body and the medium. When the perturber is introduced, a spherical contact discontinuity is excited and grows in radius at the speed of sound $\mathcal{R}_{1} = c_\mathrm{s}(t-t_-)$ which is visible in all panels of Fig.~\ref{fig:DensityWakes}. The second mechanism is sourced by significant accelerations experienced by the perturber during the pericenter approach. This second, smaller bubble of radius $\mathcal{R}_{2} \approx c_\mathrm{s}t$ is present in panels $a_2, b_2, c_2$ of Fig.~\ref{fig:DensityWakes}.\\

\setcounter{figure}{5}
\begin{figure*}
    \centering
    \includegraphics[width=\linewidth]{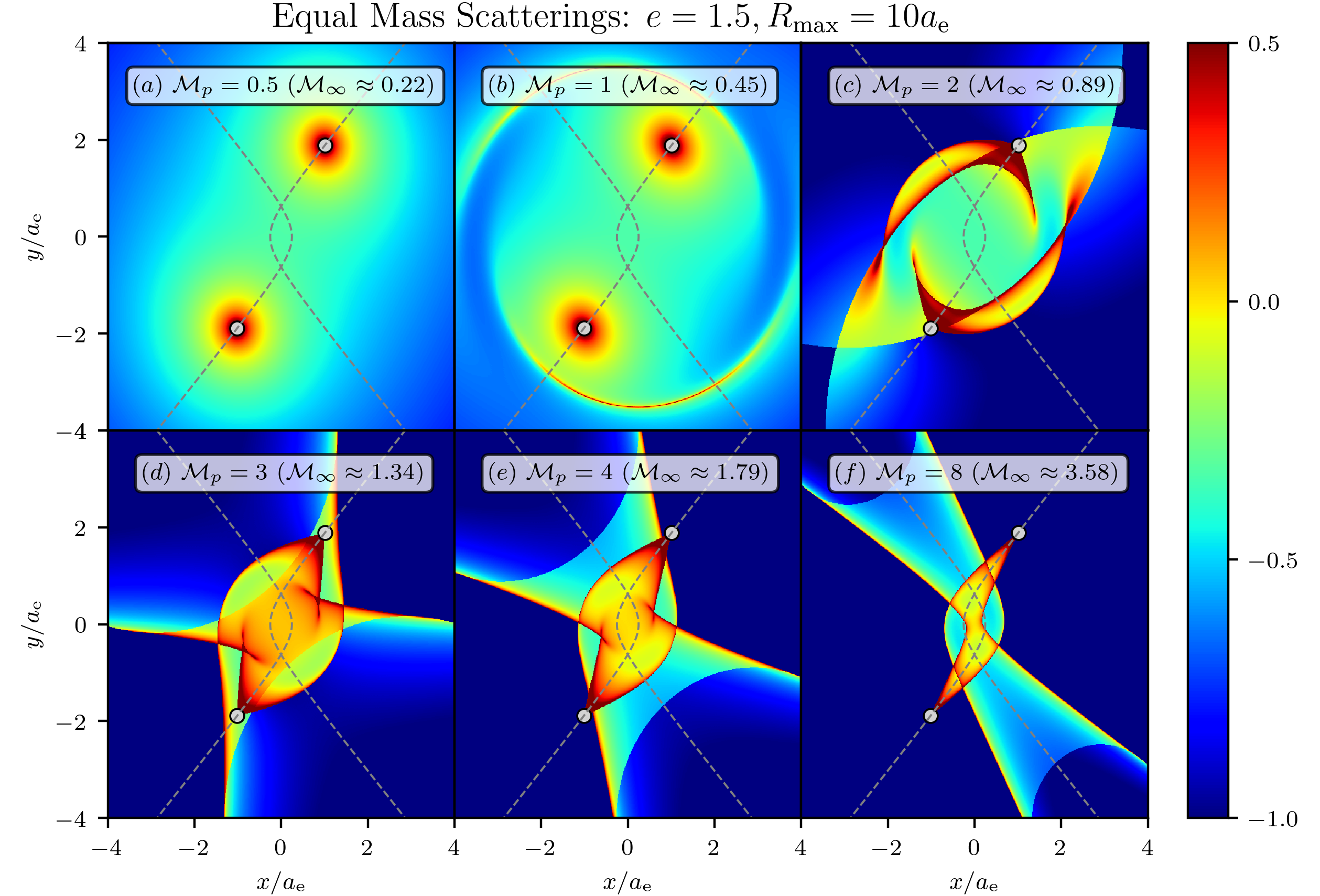}
    \caption{Logarithmic plots of the density wake $\mathcal{D}_e(\Omega t, \mathbf{x}/a_\mathrm{e},e,\mathcal{M}_p)$ created by an equal mass hyperbolic scattering. We present a selection of Mach numbers (see labels) with fixed eccentricities of $e = 1.5$ all at equal times following pericenter $t = 0.5~[2\pi\Omega^{-1}]$. The dashed gray lines delineate the fixed hyperbolic trajectories. The resolution is the same as Fig.~\ref{fig:DensityWakes} although plotted over a smaller Cartesian domain of width $8a_\mathrm{e}$.}
    \label{fig:EqualMassScatter}
\end{figure*}

Following the passage of pericenter, strictly supersonic trajectories create an ``outgoing" Mach cone (panel $c_2$ in Fig.~\ref{fig:DensityWakes}) which is orientated by an angle $\Theta$
\begin{equation}
  \Theta = 2\sin^{-1}(1 / e) = \pi - 2 A, 
\end{equation}
with respect to the initial ``incoming" cone (panel $c_1$ in Fig.~\ref{fig:DensityWakes}). The orientation angle $\Theta$ is simply the deflection angle experienced by a trajectory in hyperbolic motion (see also Fig.~\ref{fig:Classifications}). Consider the two legs which comprise each Mach cone -- one is interior to the orbit (left of the dashed white lines in panels $b_1, c_1$), and one is exterior to the orbit (right of the dashed white lines in panels $b_1, c_1$). Depending on the orbital classification, the interior leg of the incoming Mach cone can intersect with the outgoing Mach cone leading to the formation of a high-density cusp or vertex with a pinching angle $\xi$. This intersection can happen in one of two ways: For self-embedded trajectories, the intersection occurs with the exterior leg of the outgoing Mach cone at an angle of $\xi = 2(A+C)$, while for self-extracted trajectories the intersection occurs with the interior leg of the outgoing Mach cone at an angle $\xi = 2A$. These different intersections are illustrated in Fig.~\ref{fig:Pinch}, with the pinching angle $\xi$ depicted in both panels.

\setcounter{figure}{6}
\begin{figure}[H]
    \centering
    \includegraphics[width=\linewidth]{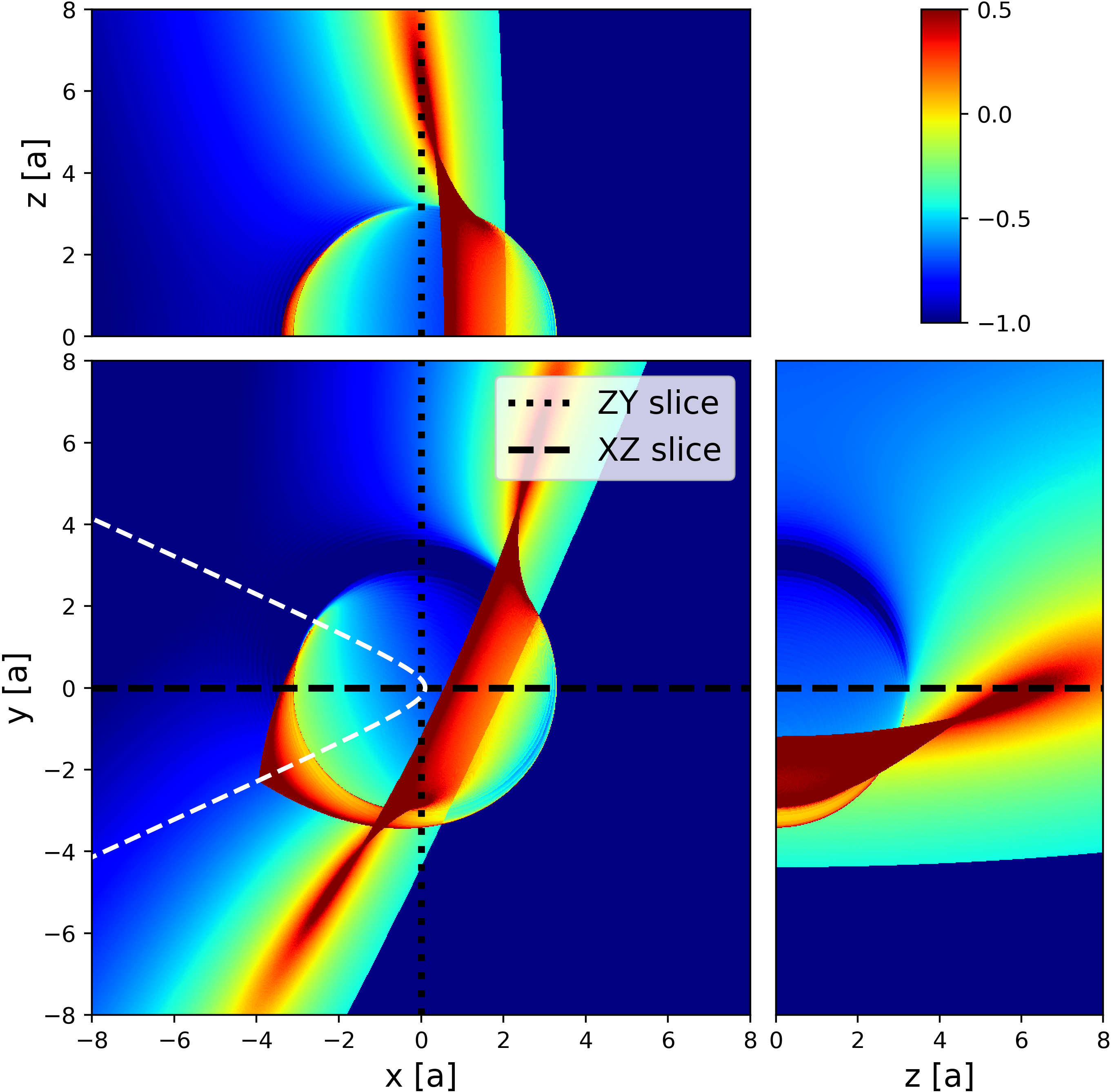}
    \caption{2D slices of the density wake created by a transonic trajectory ($e = 1.1, \mathcal{M}_p = 4.5$) across three orthogonal planes. The dashed white line denotes the orbital trajectory while the black dotted and dashed lines are the orthogonal planes plotted adjacently. 
    }
    \label{fig:3DStructure}
\end{figure}

Trajectories which can be confined to a 2D plane produce density wakes mirror-symmetric with respect to that plane. For hyperbolic trajectories defined in Eq.~(\ref{TrajectoryPrescribed}), this corresponds to a symmetry $z \to - z$ in the density wake structure -- illustrated for three orthogonal 2D slices in Fig.~\ref{fig:3DStructure}. The initialisation and acceleration bubbles (with radii $\mathcal{R}_1$ and $\mathcal{R}_2$ respectively) are present in all three panels of Fig.~\ref{fig:3DStructure}, with noticeable differences in their spatial scales.\\

Transitioning from single perturber scatterings to two-body scatterings, we illustrate the gas density morphology created by an equal-mass encounter in Fig.~\ref{fig:EqualMassScatter}. The upper row depicts three asymptotically subsonic trajectories, of which $(a)$ is strictly subsonic while $(b)$ and $(c)$ are transonic. The lower row depicts three strictly supersonic trajectories, of which $(d)$ is companion-embedded while $(e)$ and $(f)$ are companion-extracted (where $\mathcal{M}^\triangleleft_{p} \approx 3.35$). For these equal-mass, two-body interactions, there is a central point symmetry about the barycenter as discussed in \citetalias{Paper1} due to the identical nature of two perturbers. In the following section, we discuss the resulting gravitational forces exerted by such density wakes over the course of a hyperbolic encounter.\\

\setcounter{figure}{7}
\begin{figure*}
    \centering
    \includegraphics[width=\linewidth]{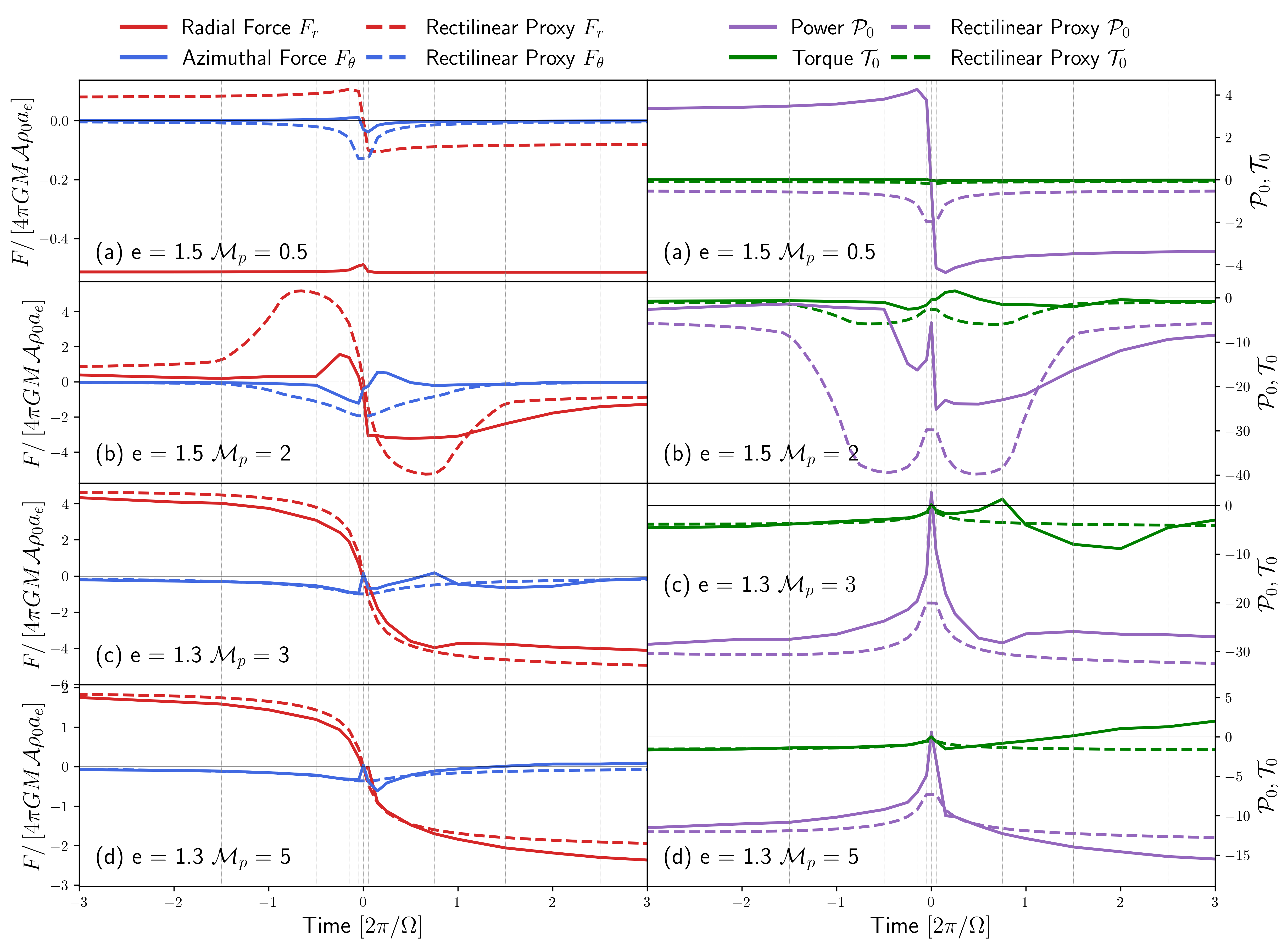}
    \caption{Left: Radial and azimuthal forces ($F_r$ in red and $F_\theta$ in blue) as a function of time. Right: Dimensionless power and torque profiles ($\mathcal{P}_0$ in purple and $\mathcal{T}_0$ in green) as defined from Eqs.~\ref{DimlessPower}, \ref{DimlessTorque}. Our selection of orbital parameters correspond to trajectories which are (a) subsonic, (b) transonic, (c) both embedded and (d) both extracted with all being equal mass scatterings over a medium of radius $R_\mathrm{max} = 100 a_e$. The dashed lines correspond to the rectilinear proxy (see \ref{sec:RectlinearProxy}).
    }
    \label{fig:ForceTime}
\end{figure*}

\subsection{Dynamical Force Profile}
\label{sec:ForceProfile}
The intricate structure of the density wakes in Figs.~\ref{fig:DensityWakes}--\ref{fig:EqualMassScatter} can be attributed to the Keplerian nature of the trajectory. The acceleration history is encoded within the density wakes, which in turn provide a time-dependent force, which we present over a fixed time window in Fig.~\ref{fig:ForceTime} for several illustrative cases.\\

\setcounter{figure}{8}
\begin{figure*}
    \centering
    \includegraphics[width=\linewidth]{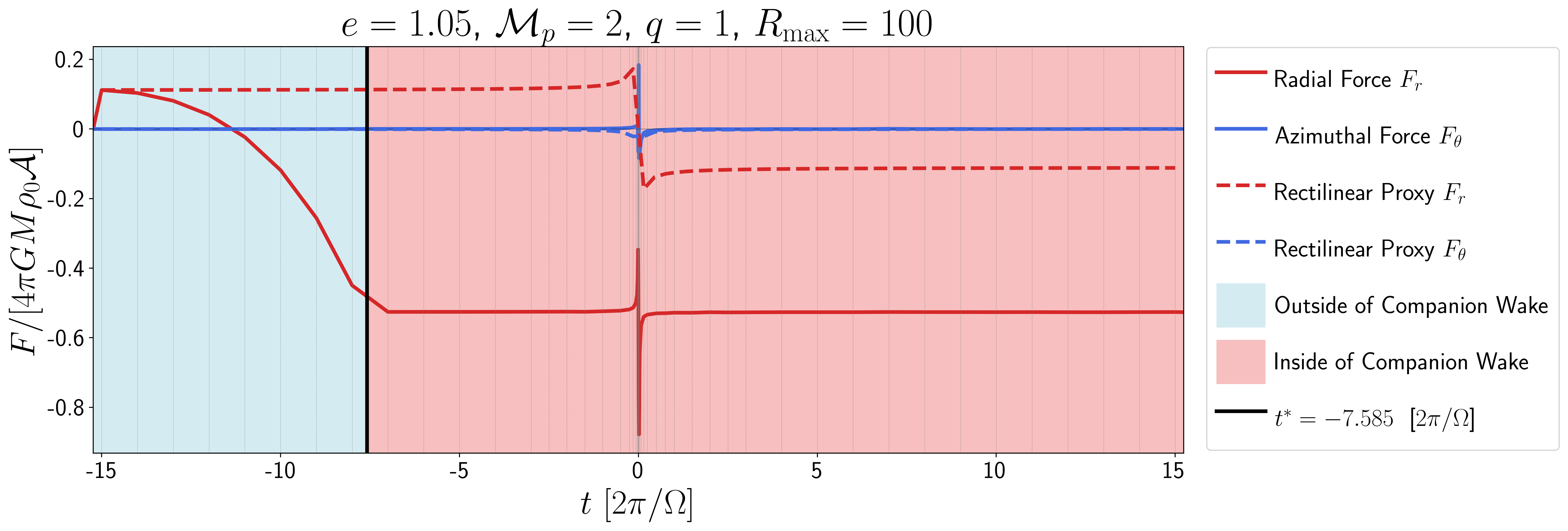}
    \caption{The time dependent force profile of an asymptotically subsonic trajectory. Here the red curves denote the radial force $F_r$, while the blue curves denote the azimuthal force $F_\theta$ (which are largely subdominant). For comparison, the dashed lines are the predictions made by the rectilinear proxy (see Section~\ref{sec:RectlinearProxy}). There is a critical time $t^*$ defined to be the moment in which a sound wave which was emitted by one perturber reaches the other perturber. \emph{Blue shade:} Early times $t<t^*$ when each perturber is outside of the wake created by its companion. \emph{Red shade:} Later times $t>t^*$ when each perturber has entered the wake created by the companion.}
    \label{fig:LongTimeDependence}
\end{figure*}

The degree of time symmetry about pericenter ($t\to -t$) in Fig.~\ref{fig:ForceTime} is determined by the amount of `memory' which has been retained by the wake. The sound speed determines the medium's ability to propagate information. Subsonic scatterings -- associated with larger sound speeds, allow the wake to ``flush out'' structure which was created during early times, rendering a density wake which is less sensitive to the trajectory's history. Consequently, panel $(a)$ in Fig.~\ref{fig:ForceTime} exhibits the most symmetry about pericenter, which corresponds to the largest sound speed of all panels. In contrast, the transonic panel (b) displays a radial force profile starkly different before and after pericenter. This discrepancy conveys the fact that the force experienced during a crossing of the sound barrier relies on whether the acceleration is positive or negative. As described in \citetalias{Paper1}, decelerating across the sound barrier can be accompanied by positive power and torque due to the advancing contact discontinuity which leads the perturber in motion (see panel $a_2$ of Fig.~\ref{fig:DensityWakes}) -- something which is not true for positive accelerations across the sound-barrier. Additionally, in panel (c) there is a `bump' in the radial force profile at $t = 0.75 [2\pi\Omega^{-1}]$ which approximately corresponds to the time when each perturber encounters the wake sourced by its companion. This interaction further contributes to asymmetries in the force profile. By contrast, the rectilinear proxy depends only on the Mach number at a given point and, therefore, the force profile remains symmetric about pericenter.\\

A strictly frictional force would lead the perturbers to lose power to the gaseous medium at all times during the encounter. Instead, panel (a) of Fig.~\ref{fig:ForceTime} illustrates a \emph{positive} power gained by the perturbers during their pericenter approach ($t<0$). In this regard, the force is not strictly frictional, but can lead to
\emph{attraction} between the two perturbers. The density wake structure in this instance is illustrated in the upper left panel of Fig.~\ref{fig:EqualMassScatter}, which strongly resembles that of two rectilinear, subsonic perturbers \citepalias[see Fig.~1 of ][]{O99}. As described by \citetalias{O99}, the force contribution (from it's own wake) is strongly diminished for subsonic motion with large cancellations of the force due to a symmetric wake structure. However, for two perturbers, the wake sourced by the companion does not exhibit such symmetry. Instead, it offers a significant force in the direction of the companion (ie., radially), resulting in positive power being gained by the perturbers. We note that this is in contrast to the rectilinear proxy, which assumes that both perturbers only experience the force of their own wake, and, are thus \emph{relentlessly frictional}. Therefore the existence of two massive perturbers as opposed to only one significantly alters the dynamical force profile and is independent of the wake memory as previously discussed.\\

Correspondingly, the torque exerted by the gas is not always negative as would be expected by the rectilinear proxy. In panels (b), (c) and (d) of Fig.~\ref{fig:ForceTime}, negative torques always occur during the pericenter approach, although not always afterwards. We highlight two mechanisms capable of generating a positive torque on the system: 
\begin{itemize}
    \item For low eccentricities, the sharp deflection in the orbital path near pericenter can occur faster than the wake can respond. This means that a torque that was previously negative on the pericenter approach can momentarily become positive during the pericenter exit (see panels $b$ and $c$ of Fig.~\ref{fig:ForceTime} and Fig.~14 of \citetalias{Paper1}). 
    \item Trajectories with orbital Mach numbers $\mathcal{M}_p \approx \mathcal{M}^\triangleright_{p}~\mathrm{or}~\mathcal{M}_p\approx\mathcal{M}^\triangleleft_{p}$ spend prolonged periods of time co-moving with either their own incoming Mach cone, or their companion's incoming Mach cone. In either case, the close proximity of this highly-dense gas can exert persistent torques on the orbit, which act over long timescales (see panel $d$ for Fig.~\ref{fig:ForceTime}).
\end{itemize}
These mechanisms for generating positive gas-induced torques enable momentum which had been previously lost to the medium to be returned to the massive perturbers helping to mitigate orbital decay.\\

\setcounter{figure}{9}
\begin{figure*}
    \centering
    \vspace{-1mm}
    \includegraphics[width=\textwidth]{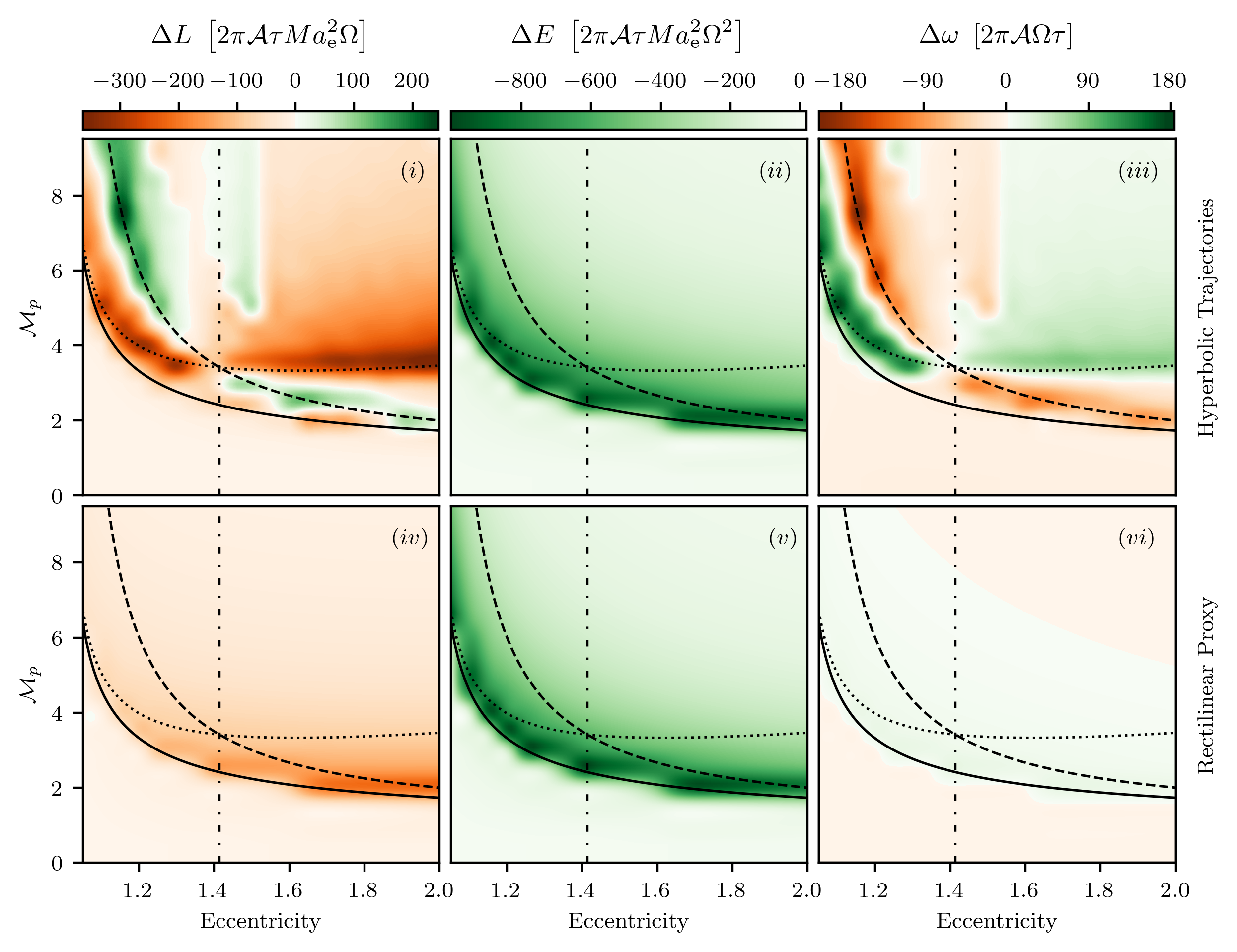}
    \caption{The change in specific energy $E$, specific angular momentum $L$ and the angle of precession $\omega$ as a function of orbital eccentricity $e$ and pericenter Mach number $\mathcal{M}_p$. The dashed black line separates strictly supersonic (above) from transonic (below) and the dotted black line separates extracted orbits (above) from embedded ones (below). Our data is interpolated over a grid of spacing $0.05$ in eccentricity and $0.5$ in Mach number.}
    \label{fig:OrbitalElementChanges}
\end{figure*}

The timescale required for the first sound-wave to propagate from one perturber to the other represents a critical timescale $t^*$ for asymptotically subsonic trajectories -- physically corresponding to the moment when each perturber enters the wake sourced by its companion. We can make a simple analytical estimate for this timescale by equating the initial separation of the perturbers with the distance travelled by of one of the bodies at time $t$, plus the distance travelled by a sound wave, $c_\mathrm{s}(t - t_-)$. To facilitate obtaining a simple analytical result, we assume that the radius of the medium is large $R_\mathrm{max} \gg a_\mathrm{e}$ so that both perturbers move at approximately constant speed and their initial separation is $d \approx 2\sqrt{1 + e^2\sinh^2\sigma(t_-)}$. The critical time $t^*$ when a sound wave sourced from one perturber arrives at the companion thus occurs at time
\begin{equation}
    t^* \approx t_- + \frac{2 \sqrt{1+e^2\sinh^2{\sigma(t_-)}}}{c_\mathrm{s} (1 + \mathcal{M}_\infty)}.
    \label{Tcrit}
\end{equation}
In Fig.~\ref{fig:LongTimeDependence} we illustrate an example of the time dependent force profile for an asymptotically subsonic trajectory. At very early times $t\ll t^*$, (although after a very short initial transient) the force exerted by the gas agrees excellently with the value predicted by \citetalias{O99} (around $t \approx -15 [2\pi/\Omega]$). During these early times, each perturber is predominantly influenced by its own wake. Over time, the companion wake grows and significantly changes the force profile. First, the sign of the radial force changes (around $t \approx -11.5 [2\pi/\Omega]$) before saturating around $t = t^*$. {This negative radial force leads to positive power being transmitted to the perturbers by the gaseous medium.}
In general, we observe this attraction for all asymptotically subsonic trajectories for times $t > t^*$.

\subsection{Orbital Dynamics}
\label{sec:OrbitalDynamics}
The power and torque exerted by the gas will lead to changes in energy $\Delta{E}$, angular momentum $\Delta{L}$, and a gas-induced precession angle $\Delta{\omega}$. In Fig.~\ref{fig:OrbitalElementChanges} we present the changes in these orbital parameters across a parameter space spanned by eccentricity and pericenter Mach number for equal mass perturbers and a fiducial value for $R_\mathrm{max} = 100 a_\mathrm{e}$. We compare our results to the rectilinear proxy (bottom row) as introduced in Section \ref{sec:RectlinearProxy}.\\

In Fig.~\ref{fig:OrbitalElementChanges}, the change in orbital angular momentum (upper left panel) can be either positive (green) or negative (orange). For a large medium, i.e., $R_\mathrm{max} \gg a_\mathrm{e}$, the majority of the change in orbital elements arises from the asymptotic regime, in which one may feel tempted to assume that employing the rectilinear proxy would be appropriate. In fact, visible from the lower row, this is not true as the proxy predicts a strict loss of angular momentum across parameter space. This shows that the torques exerted on the perturbers arise from more intricate features of the gas dynamics which is not captured by the rectilinear proxy. From the upper left panel of Fig.~\ref{fig:OrbitalElementChanges}, maximal angular momentum gain occurs near the self-extraction separatrix ($\mathcal{M}^\triangleleft_{p}(e)$, black dashed line) and maximal momentum loss occurs near the companion-extraction separatrix ($\mathcal{M}^\triangleright_{p}(e)$, black dotted line) as orbits falling along these lines are co-moving with a previously generated Mach cone. This suggests that the self-wake generated by each perturber is responsible for exerting strong positive torques, whereas the companion-wake is responsible for providing strong negative torques over the course of the encounter.\\

The change in orbital energy is always negative across the parameter space given in Fig.~\ref{fig:OrbitalElementChanges}. As expected, this demonstrates that the medium is a dissipative source of orbital energy, which occurs maximally near the transonic line (solid black in all panels). Contrary to the change in orbital angular momentum, the change in energy is qualitatively similar to the results of the rectilinear proxy. This agreement suggests that the asymptotic Mach number $\mathcal{M}_\infty$ is the only important parameter when calculating the change in energy. When $\mathcal{M}_\infty\approx 1$, each perturber experiences the maximum force (as described by the \citetalias{O99}) over a prolonged period of time which results in maximal energy loss. This is analogous to the findings of \citetalias{Paper1} wherein a semi-major axis decay is found to occur predominantly at the transonic separatrix -- where the mean Mach number most closely approaches unity. In summary, the change in orbital energy is only sensitive to the asymptotic Mach number, whereas the change in orbital angular momentum is sensitive to more intricate features in the density wake structure.\\

\begin{figure*}
    \centering
    \includegraphics[width=\linewidth]{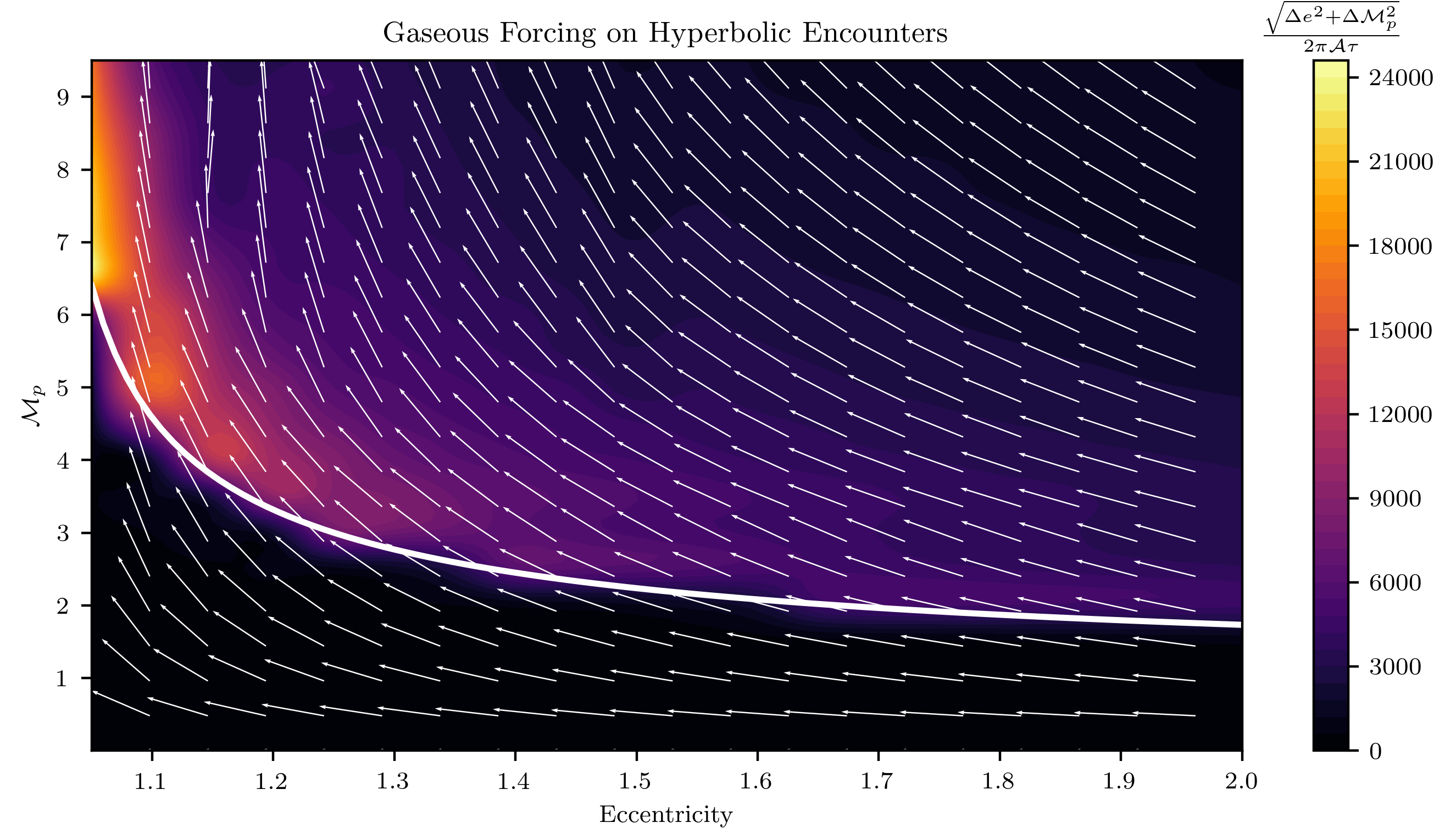}
    \caption{A flow diagram of the orbital parameters (eccentricity $e$ and pericenter Mach number $\mathcal{M}_p$) for equal-mass trajectories embedded in media of radius $R_\mathrm{max} = 100a_e$. The white arrows depict the normalised unit vectors of direction, while the colorbar indicates the rate of orbital evolution. The white solid line indicates the transonic separatrix $\mathcal{M}_\infty = 1$.}
    \label{fig:FlowPlot}
\end{figure*}

The gas-induced precession angle $\Delta\omega$ (top right of Fig.~\ref{fig:OrbitalElementChanges}) displays much of the same structure as the change in orbital angular momentum, albeit with a relative difference in the sign. Judging by the maximal and minimal values of $\Delta\omega$, the torque exerted by the self-wake drives a negative precession, whereas the companion wake drives a positive precession. Once again, these features are not captured by the rectilinear proxy (bottom right panel of Fig.~\ref{fig:OrbitalElementChanges}). In fact, while the rectilinear proxy does predict a non-zero precession rate, it is often orders of magnitude subdominant to that found by employing consistent hyperbolic trajectories. To understand this discrepancy, we note that the rectilinear proxy predicts an (approximately\footnote{Perfectly odd in the case subsonic motion, while supersonic motion grows in magnitude logarithmically with time (Eq.15 of \citetalias{O99}). Therefore the force experienced post-pericenter passage is greater for supersonic motion than the force experienced during the pericenter approach.}) odd radial force $F_r(-t) \approx -F_r(t)$ and even azimuthal force
$F_{\theta}(-t) \approx F_{\theta}(t)$. When these forces are weighted with the appropriate sine and cosine terms in Eq.~(\ref{Precession}), the rate of precession becomes an (approximately) odd function of time, $\dot{\omega}(-t) \approx -\dot{\omega}(t)$. Hence, when integrated over the encounter, large cancellations occur between the pericenter approach and the pericenter exit. This cancellation results in a weak gas-induced precession angle $\Delta \omega$. Conversely, there is a noticeable lack of symmetry about pericenter in the force profiles of Fig.~\ref{fig:ForceTime} (as previously discussed in Section~\ref{sec:ForceProfile}) 
which prevents significant cancellations of $\Delta\omega$, 
thus generating a non-zero gas-induced precession angle, as observed in Fig.~\ref{fig:OrbitalElementChanges}.

\subsection{Changes to Orbital Parameters}
\label{sec:OrbitalEvolution}
We seek to translate changes in the orbital elements ($\Delta E$, $\Delta L$) to changes in the orbital parameters ($\Delta e, \Delta \mathcal{M}_p$). First, we express the eccentricity in terms of the orbital energy and angular momentum,
\begin{equation}
    e = \sqrt{1+\frac{2EL^2}{G^2M^4}}.
\end{equation}
Next, we compute the variation to eccentricity $\Delta{e}$ given changes to energy $\Delta E$ and angular momentum $\Delta L$,

\begin{equation}
    \Delta e = \frac{e^2 - 1}{e}\left(\frac{a\Delta E}{GM^2} + \frac{a\Omega\Delta L}{GM^2\sqrt{e^2-1}}\right).
    \label{Delta_e}
\end{equation}
Similarly, the change in pericenter Mach number (Eq.~\ref{HyperbolicVelocity}) can be expressed in terms of $\Delta E$ and $\Delta e$ as,
\begin{equation}
    \Delta\mathcal{M}_p = -2 \mathcal{M}_p \frac{\Delta E}{Ma^2\Omega^2} - \mathcal{M}_p\frac{\Delta{e}}{e^2 - 1}.
    \label{Delta_Mp}
\end{equation}
Finally, given the variations to eccentricity and pericenter Mach number in Eqs.~(\ref{Delta_e}), (\ref{Delta_Mp}) we construct the orbital evolution vector field
\begin{equation}
    \mathbf{f}(e,\mathcal{M}_p) = \Delta{e}\hat{e} + \Delta{\mathcal{M}}_p\hat{\mathcal{M}}.
    \label{VectorEvolution}
\end{equation} 
In Fig.~\ref{fig:FlowPlot} we plot $\mathbf{f}$ across our parameter space for a fiducial value of $R_\mathrm{max}=100a_\mathrm{e}$.\\

The gaseous medium tends to damp orbital eccentricity, while increasing the pericenter Mach number in Fig.~\ref{fig:FlowPlot}. Contrary to the case for bound systems (Fig.12 of \citetalias{Paper1}), a single point in parameter space cannot be interpreted as tracing a track through the parameter space as the orbit is not closed, meaning there is not a cyclic nature over which the orbit can evolve on secular timescales. As an alternative interpretation, Fig.~\ref{fig:FlowPlot} illustrates that an ensemble of hyperbolic orbits (with varying initial orbital parameters) will become more eccentric ($e \to 1$) and supersonic as a result of the interaction with the homogeneous gaseous medium. This is reminscent of the findings in \citetalias{Paper1}, in which bound orbits are found to evolve into a highly eccentric ($e \to 1$) supersonic state.
\subsection{Dependence on $R_\mathrm{max}$}
\label{sec:DependenceOnRmax}
In Fig.~\ref{fig:RadialScaling} we vary the size of the gaseous medium $R_\mathrm{max}$, for $5$ trajectories of fixed orbital parameters. We choose these parameters to be representative samples of transonic, supersonic, self-extracted and self-embedded regions of parameters space.\\

A simple analytical fit can be obtained for the energy dissipation based on the analytical model of \citetalias{O99} (top panel of Fig.~\ref{fig:RadialScaling}). In the limit of a large medium $R_\mathrm{max}\gg 1$, we assume that the majority of energy is dissipated during the asymptotic regime, in which the tangential force is well modelled by $\mathbf{F}_\mathrm{O99}$ (Eqs.~(12)-(15) of \citetalias{O99}). The energy dissipated is therefore, 
\begin{equation}
    \Delta{E} = \int_{t_-}^{t_+} dt ~2\mathbf{F}_\mathrm{O99}\cdot{\mathbf{v}_\infty},
    \label{IntegratePowerO99}
\end{equation}
where $t_+ = R_\mathrm{max}/v_\infty$, $t_- = - R_\mathrm{max}/v_\infty$ and $v_\infty = a_\mathrm{e}\Omega/2$. Performing the integral in Eq.~(\ref{IntegratePowerO99}) yields the change in orbital energy
\begin{equation}
    \Delta{E} = 2\pi \mathcal{A}\tau M a_e^2\Omega^2 \mathcal{E}(R_\mathrm{max}, \mathcal{M}_\infty),
\end{equation}

\begin{widetext}
\begin{equation}
    \mathcal{E}(R_\mathrm{max}, \mathcal{M}_\infty) = -K \begin{cases} 
          \frac{2R_\mathrm{max}}{a_e}\frac{1}{\mathcal{M}_\infty^2}\left[\frac{1}{2}\ln\left(\frac{1+\mathcal{M}_\infty}{1-\mathcal{M}_\infty}\right) - \mathcal{M}_\infty\right] & \mathrm{if}~\mathcal{M}_\infty < 1 \\
          \frac{2R_\mathrm{max}}{a_e}\frac{1}{\mathcal{M}_\infty^2}\left[\frac{1}{2}\ln\left(1-\frac{1}{\mathcal{M}_\infty^2}\right) -1 + \ln\left(\frac{2R_\mathrm{max}}{r_\mathrm{min}}\right)\right] & \mathrm{if}~ \mathcal{M}_\infty > 1 .
          \end{cases}
          \label{EnergyFit}
\end{equation}
\end{widetext}
The parameter $K$ has been introduced as a fitting factor, which we determine to be $K=0.9775$ by requiring it to minimise the geometric mean of the fractional error between our results and Eq.~(\ref{EnergyFit}). In the top panel of Fig.~\ref{fig:RadialScaling}, we illustrate this analytical model along with our results for $\Delta{E}(R_\mathrm{max})$. We note that the force exerted by the medium grows logarithmically in time, although the change in energy scales superlinearly with $R_\mathrm{max}$. The accuracy of this model is perhaps unsurprising, as the energy dissipation depends primarily on the asymptotic Mach number (as previously discussed).\\

\setcounter{figure}{12}
\begin{figure*}
    \centering
\includegraphics[width=\linewidth]{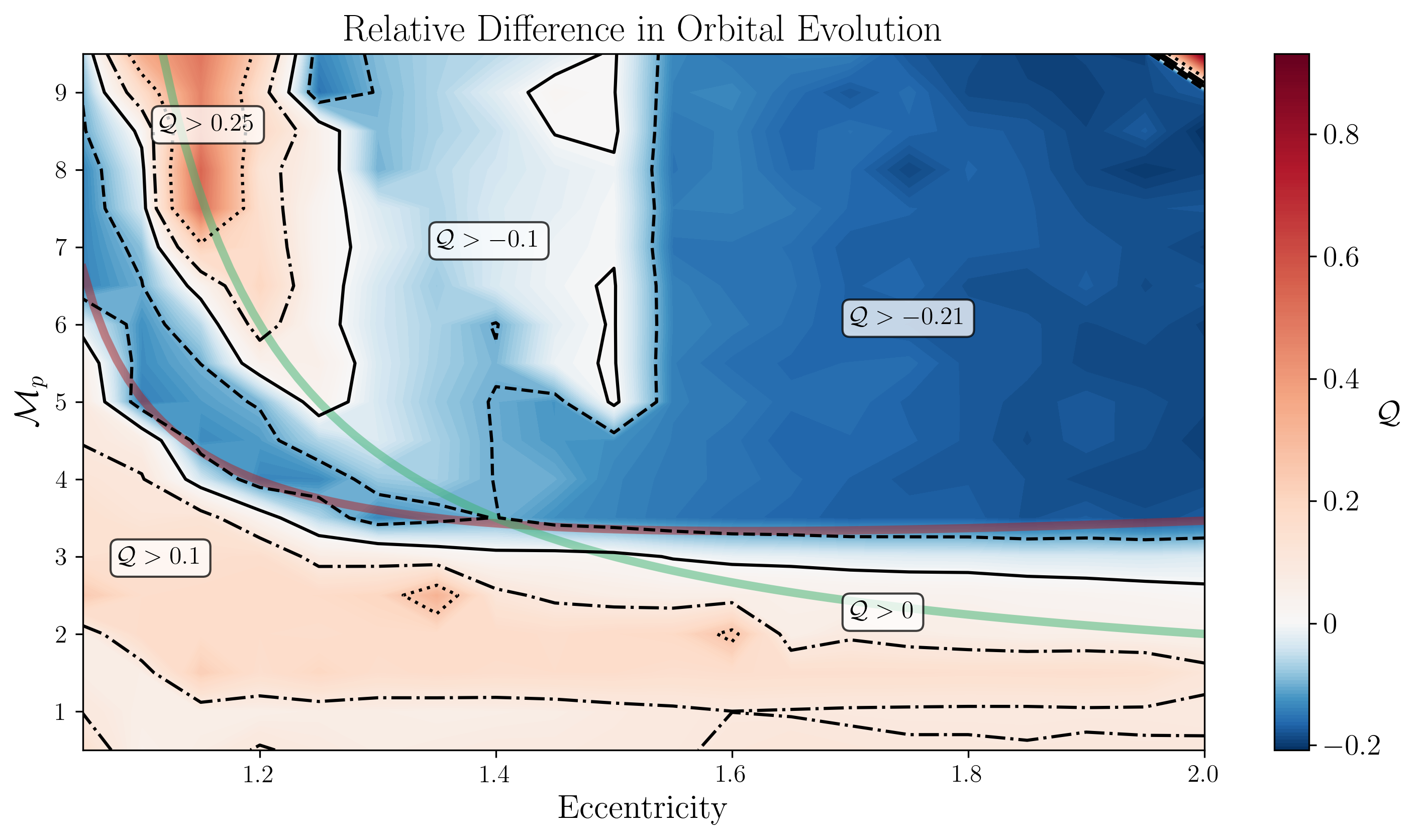}
    \caption{The normalised alignment coefficient $\mathcal{Q}$ (see Eq.~\ref{AligmentCoefficient}) used to compare the orbital evolution results between consistent hyperbolic trajectories and the rectilinear proxy. The red solid curve indicates the companion-embedding criterion $\mathcal{M}_p^\triangleleft(e)$, while the green curve represents the self-embedding criterion $\mathcal{M}_p^\triangleright(e)$. The black dashed, solid, dash-dotted and dotted lines are isocontours of $\mathcal{Q} = [ -0.1, 0, 0.1, 0.25]$ respectively. We adopt a medium scale of $R_\mathrm{max} = 100 a_e$ in the figure above.}
    \label{fig:VectorDifference}
\end{figure*}

\setcounter{figure}{11}
\begin{figure}[H]
    \centering
    \includegraphics[width=\linewidth]{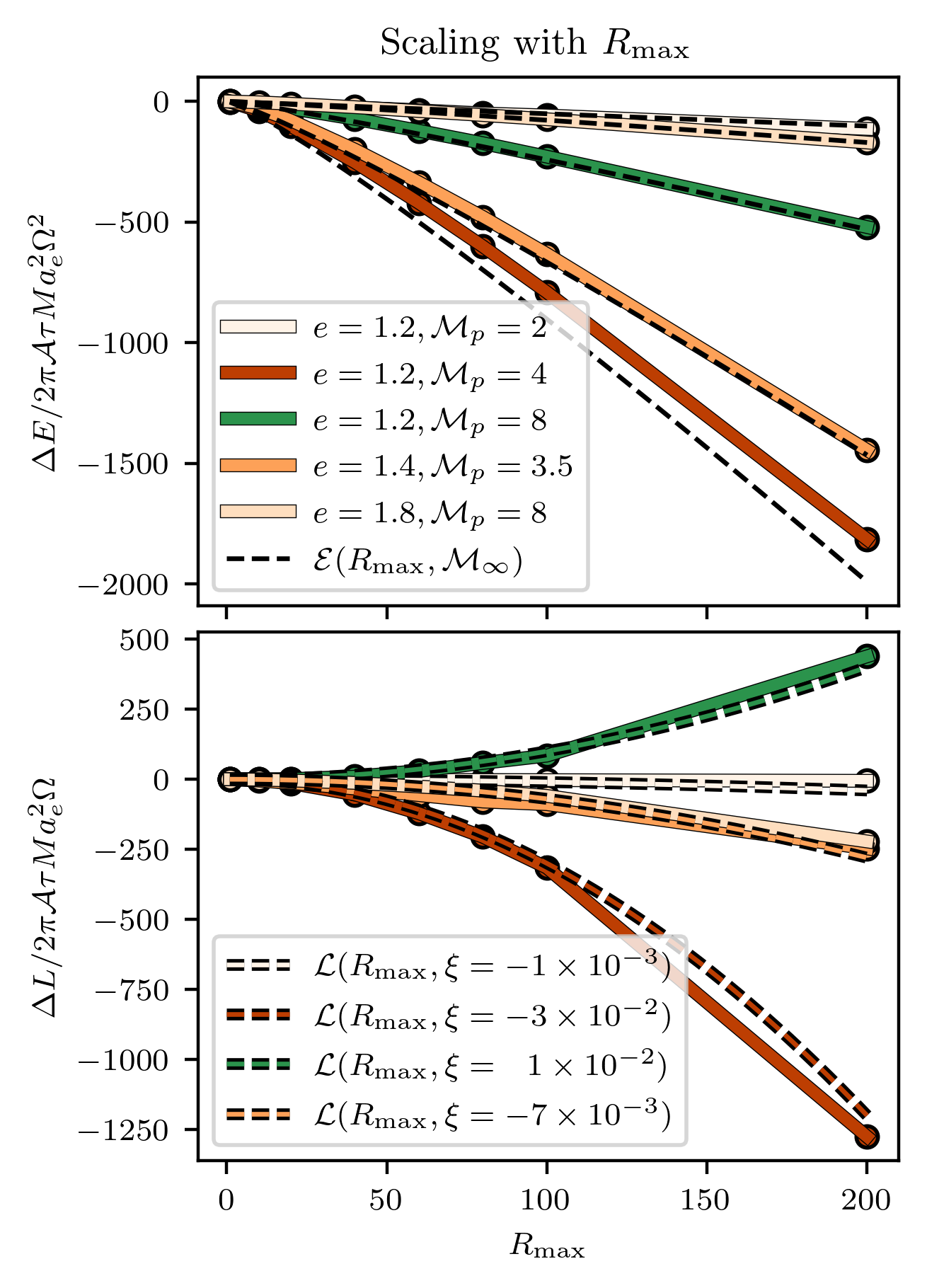}
    \caption{Radial dependence of the change in energy $\Delta{E}$ and angular momentum $\Delta{L}$ for a set of fixed trajectories (see legend).}
    \label{fig:RadialScaling}
\end{figure}

In contrast, the change in orbital angular momentum is dominated by the torques exerted due to wake intersections and accelerations at pericenter -- not just the asymptotic Mach number, proving far more challenging to accurately model across the parameter space. Instead, in the lower panel of Fig.~\ref{fig:RadialScaling} we present a series of phenomenological power-law fits of the form
\begin{equation}
    \Delta L = 2\pi \mathcal{A}\tau Ma_e^2\Omega \mathcal{L}(R_\mathrm{max}, \xi),
\end{equation} 
where 
\begin{equation}
    \mathcal{L}(R_\mathrm{max}, \xi) = \xi\left({R_\mathrm{max}}/{a_\mathrm{e}}\right)^2.
    \label{MomentumFit}
\end{equation}
The accuracy of these fitting functions in Fig.~\ref{fig:RadialScaling} suggests that the change in orbital angular momentum typically scales quadratically with the size of the medium. We provide the numerical values of $\xi$ in the legend of Fig.~\ref{fig:RadialScaling}.

\subsection{Comparison to the Rectilinear Proxy}
In this section, we compare our findings for orbital evolution (see Fig.~\ref{fig:FlowPlot}) to the results obtained by using the rectilinear proxy. Following Eq.~(\ref{VectorEvolution}), we begin by defining the vector field flow of the rectilinear proxy as \begin{equation}
    \mathbf{g}(e,\mathcal{M}_p) = \tilde{\Delta} e \hat{e} + \tilde{\Delta} \mathcal{M}_p\hat{M}_p.
\end{equation}
To quantify how this vector field differs from the results presented in Fig.~\ref{fig:FlowPlot}, we introduce a normalised alignment coefficient
\begin{equation}
    \mathcal{Q} = \frac{\mathbf{f}\cdot\mathbf{g}}{|\mathbf{f}|^2} - 1,
    \label{AligmentCoefficient}
\end{equation}
representing the fractional difference of $\mathbf{g}$ with respect to $\mathbf{f}$ (which is defined in Eq.~\ref{VectorEvolution}). We always find that $\mathbf{f}\cdot{\mathbf{g}}>0$, suggesting the direction of orbital evolution is in overall agreement. We thus identify positive $\mathcal{Q}$ regions as those in which the rectilinear proxy predicts a faster orbital evolution than our findings, and likewise, a negative $\mathcal{Q}$ value corresponds to regions in paramter space where the rectilinear proxy predicts a orbital evolution slower than our findings. We illustrate the normalised alignment coefficient in Fig.~\ref{fig:VectorDifference} for a fiducial value of $R_\mathrm{max} = 100a_e$.\\

There exists a narrow corridor of negative $\mathcal{Q}$ values at low eccentricities in Fig.~\ref{fig:VectorDifference}. Within this corridor is the red solid curve, representing the companion-embedding sepratrix. Along this seperatrix occurs a maximal angular momentum loss (see Fig.~\ref{fig:OrbitalElementChanges}) producing a rapid orbital evolution which is not captured by the rectilinear proxy, thus leading to a negative $\mathcal{Q}$ value. Correspondingly, above this corridor exists a positive $\mathcal{Q}$ peninsula, traced neatly by the self-embedding seperatrix (green curve). Orbits along this line gain orbital angular momentum from the wake they generated during the pericenter approach, helping to mitigate orbital decay. Therefore, the orbital evolution  is slower than the rectilinear proxy, leading to a positive $\mathcal{Q}$ value. Hence, the orbital classifications introduced in Section~\ref{sec:Classifications} provides a natural framework to understand the regions in which there is a maximal difference between our results and the rectilinear proxy. Across this parameter space, the rectilinear proxy is typically within $25\%$ of our findings (ie. all regions outside of the dotted contours in Fig.~\ref{fig:VectorDifference}). \\ 

Next, we quantify the normalised alignment coefficient $\mathcal{Q}$ for different sized media in Fig.~\ref{fig:QRmaxScaling}. By choosing $5$ characteristic orbits (the same as in Fig.~\ref{fig:QRmaxScaling}) we measure $\mathcal{Q}$ as a function of $R_\mathrm{max}$. \\

For small media scales ($R_\mathrm{max}\leq 60 a_\mathrm{e}$), the rectilinear proxy does not accurately capture the orbital evolution of equal-mass hyperbolic perturbers (see Fig.~\ref{fig:QRmaxScaling}). The normalised alignment coefficient $|\mathcal{Q}|\sim 1$ signifies differences of order unity between our findings and the rectilinear proxy. We attribute these differences to increased emphasis placed upon the pericenter passage for small media scales, during which time the density wake structure substantially differs from that of two rectilinear motion perturbers (see Fig.~\ref{fig:EqualMassScatter}). As a result, the rectilinear proxy is not an accurate model for the orbital evolution in small media $R_\mathrm{max}\leq 20 a_\mathrm{e}$.\\

\setcounter{figure}{13}
\begin{figure}[H]
    \centering
    \includegraphics[width=\linewidth]{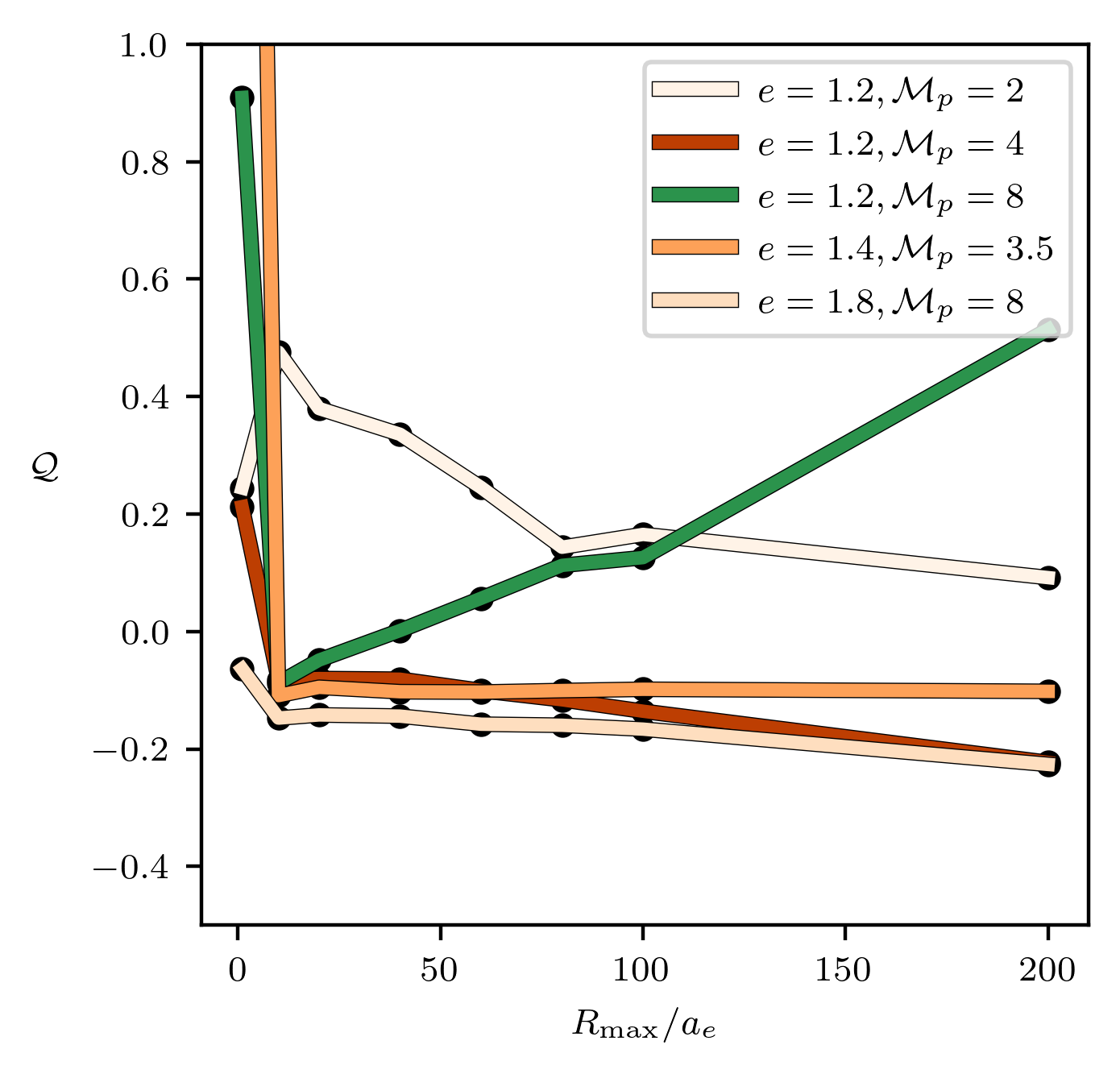}
    \caption{The normalised alignment coefficient $\mathcal{Q}$ (see Eq.~\ref{AligmentCoefficient}) for a selection of orbital parameters (coloured lines) as a function of $R_\mathrm{max}$.}
    \label{fig:QRmaxScaling}
\end{figure}

In the intermediate regime ($60 a_\mathrm{e}<R_\mathrm{max}\leq 100 a_\mathrm{e}$), there is good agreement between both models for orbital evolution. The normalised alignment coefficient $|\mathcal{Q}|\lesssim 0.25$ within this interval (see Fig.~\ref{fig:QRmaxScaling}) suggesting the rectilinear proxy predicts an orbital evolution within $25\%$ of our findings. For these medium scales, the change in energy is dominant over the change in orbital angular momentum, for which an accurate model of the energy dissipation is sufficient to capture the subsequent orbital evolution.\\

Finally, for large media ($R_\mathrm{max} >100 a_\mathrm{e}$), significant differences re-emerge between our findings and the rectilinear proxy for orbital evolution. As discussed in Section~\ref{sec:OrbitalDynamics}, one may feel tempted to assume that for very large media, the rectilinear proxy would be more appropriate; however, this is not true, evident from Fig.~\ref{fig:QRmaxScaling}. For medium radii beyond $R_\mathrm{max}=100a_\mathrm{a}$, the change in orbital angular momentum is non-negligible as it follows a stronger scaling with $R_\mathrm{max}$ than energy does (see Fig.~\ref{fig:RadialScaling}). In this regime, an accurate model of the change in orbital angular momentum is essential, which the proxy fails to provide. Thus, for large media, the rectilinear proxy predicts an orbital evolution substantially different from our findings using consistent hypebolic perturbers.\\

\subsection{Conditions for Gas-Capture}
We have not yet considered the transition from unbound hyperbolic orbits to bound elliptical ones, as our assumption of fixed orbits disregards the feedback of gas throughout the encounter. In this subsection, we briefly discuss the criteria for so-called gas captures based on the energy dissipation results for prescribed orbits.\\

As discussed in Section~\ref{sec:DependenceOnRmax} the energy dissipated by the gaseous medium can be reliably fit by the rectilinear proxy. Therefore, if we assume that this is also true for orbits experiencing live gas feedback, we can define the criteria for gas-capture. Initially, the total orbital energy $E_0$ is given by
\begin{equation}
    E_0 = \frac{1}{2}{M\left(\frac{a_\mathrm{e}\Omega}{2}\right)^2}.
\end{equation}
Given the energy dissipated during the encounter $\Delta E$ in Eq.~\ref{EnergyFit}, we can define the criterion for gas-capture as,
\begin{equation}
    \mathcal{E} \geq -\frac{1}{16\pi\mathcal{A}\tau},
\end{equation}
where we recall $\mathcal{A}=GM/a_\mathrm{e}c_\mathrm{s}^2$ and $\tau = G\rho_0/\Omega^2$. In Fig.~\ref{fig:CaptureCriteria} we illustrate the energy dissipated as a function of the asymptotic Mach number $\mathcal{M}_\infty$ and medium scale $R_\mathrm{max}$. As expected, the amount of energy dissipated increases monotonically with $R_\mathrm{max}$ and peaks at $\mathcal{M}_\infty=1$. Additionally, we overplot isocontours of $\mathcal{A}\tau$ which distinguishes between the different regions of parameter space which can (right of dashed lines) form bound systems and those which cannot (left of dashed lines) form bound systems through gas-assisted capture. A larger value of $\tau$ corresponds to a denser medium while a larger value of $\mathcal{A}$ corresponds to a perturber sourcing larger density perturbations and thus greater non-linearities. %\pagebreak

\setcounter{figure}{14}
\begin{figure}[h]
    \centering
    \includegraphics[width=\linewidth]{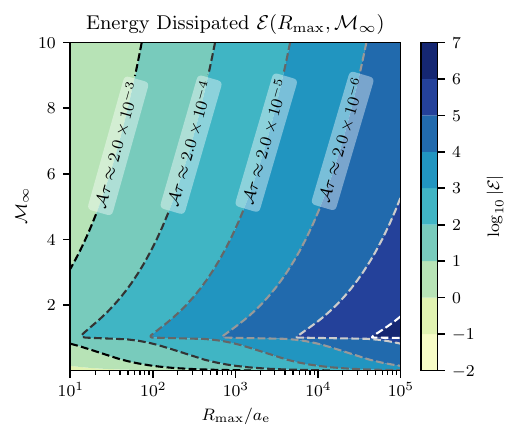}
    \caption{Energy dissipated as a function of the medium radius $R_\mathrm{max}$ and asymptotic Mach number $\mathcal{M}_\infty$. The isocontours indicate the values for $\mathcal{A}\tau$ for which capture will occur.}
    \label{fig:CaptureCriteria}
\end{figure}

\section{Discussion and Conclusion}\label{sec:Conclusion}
\subsection{Summary}
In this paper, we present a linear model describing equal-mass two-body scatterings embedded in a homogeneous medium. We discuss the wake morphology generated by the encounter, the time-dependent forces exerted onto the perturbers by the gas, the subsequent orbital evolution of the scattering and finally we compare our results to the analytical model of \citetalias{O99} via the rectilinear proxy as defined in Section~\ref{sec:RectlinearProxy}. Below, we summarise our main results:\\

\begin{enumerate}
    \item The gaseous medium tends to promote supersonic, gas-captures. We find that orbital eccentricity is damped by the gas while the pericenter Mach number always increases due to a shrinking semi-major axis (see Fig.~\ref{fig:FlowPlot}).
    \item During the pericenter approach, trajectories which are asymptotically subsonic (ie. $\mathcal{M}_\infty<1$) gain positive power due to their interaction with the gas. In spite of the name `Gaseous Dynamical Friction' the force exerted on the perturbers is not strictly frictional (see Fig.~\ref{fig:LongTimeDependence}).
    %Therefore, despite the name `Gaseous Dynamical Fricton', the force exerted on the system is not strictly frictional.
    \item Hyperbolic encounters can be classified by those which do and those which do not re-intersect the wakes which were generated during the pericenter approach (see Section~\ref{sec:Classifications}). We found that these orbital classes exhibit morphological differences in their density wake structure while also coinciding with the extremal changes to orbital energy, angular momentum and apsidal precession.
    \item The argument of periapses experiences significant precessions over the course of the scattering. This suggests that rapid angular dispersion may occur for two-body scatterings in gaseous media. 
    \item The change in orbital energy is only sensitive to the asymptotic Mach number, whereas the change in orbital angular momentum is sensitive to more intricate features in the density wake structure. This allows us to provide analytical estimates for the energy dissipated during the encounter (Eq.~\ref{EnergyFit}) based on the analytical model of \citetalias{O99}.  
    \item The rectilinear proxy predicts orbital evolution results broadly consistent with our findings, although the associated evolutionary timescales are typically shorter than those found when using consistent hyperbolic trajectories (see Fig.~\ref{fig:VectorDifference}). We find that the rectilinear proxy is most accurate for medium radii of $20a_\mathrm{a}<R_\mathrm{max}\leq100a_\mathrm{e}$, where the timescale for orbital evolution is within $\sim 25\%$ of our findings.
\end{enumerate}

Our findings for orbital evolution are in overall agreement with \cite{Paper1}, where elliptical Keplerian orbits are found to evolve towards a highly eccentric supersonic state (see their Fig.~12). Therefore, according to linear theory, the parabolic eccentricity $e = 1$ is an attractor for all point-particle perturbers on presribed Keplerian orbits embedded in a homogeneous, static medium.\\

\subsection{Additional Physics and Caveats}
\label{sec:Caveats}
The results obtained in this study are reliant on a number of simplifying assumptions, which we list below. These assumptions should be taken into consideration when applying the findings of this paper to realistic astrophysical settings.

\begin{itemize}
    \item \textbf{Fixed Trajectories:} We have neglected the live-feedback of gas on the orbit, which is a valid approximation only in the regime of low gas-density ($\tau \ll 1$). Depending on the system, this assumption may need to be relaxed to model orbits which change significantly over the course of the encounter \citep{RowanAGN, Rowan24, Whitehead2024, 2024A&A...683A.135T}.
    \item \textbf{Background Gas Dynamics:} We have assumed that the background gas is both homogeneous and isotropic. Accounting for more realistic density profiles \citep{Hirai22, Gali24} and/or velocity profiles \citep[eg.][for an accretion disk with differential rotation]{Cresswell2006,Bitsch_2010,Muto11,Fairbairn25a, Fairbairn25b} may lead to different results for orbital evolution. For a Keplerian disk in vertical hydrostatic equilibrium, the assumption of a homogeneous static medium is accurate when $L/H\ll 1$, where $H$ is the disk scale height and $L\sim R_\mathrm{max}$ is the spatial scale of the density wakes (see Appendix~\ref{ApplicationsInAstrophysicalDisks} for more detail).
    \item \textbf{Background Gas Thermodynamics:} We have modelled the medium as an inviscid, laminar fluid. However, luminous perturbers in opaque, thermally diffuse fluids have been found to experience positive accelerations rather than frictional decelerations \citep{Park_2017, Velasco19}. The effect of turbulence also decreases the efficiency of the drag force at $\mathcal{M}=1$ although increases in efficiency at larger Mach numbers \citep{Lescaudron23}.
    \item \textbf{Non-Linearities:} The linear solutions we have obtained for the density wake structure are approximations to the full non-linear solution, likely breaking down close to each perturber and when $\mathcal{M}_\infty\gg 1$ (see Section~\ref{sec:FrictionForce}). To capture these non-linear effects, numerical hydrodynamical simulations are required \citep[eg. see][]{Kim_NonLin1, Kim_NonLin2,RowanAGN, Rowan24, Whitehead2024}. Additionally, the non-linear propagation of sound waves can cause them to steepen and shock -- likely leading to the dissipation of Mach cones and contact discontinuities (see also Section~4.2 of \citealt{Paper1}).
    \item \textbf{Gas Accretion:} The accretion of material will provide an additional force on the perturber,
    as well as increasing the total effective mass of the system \citep{SanchezSalcedo2018}. \cite{Suzuguchi24} finds that the inclusion of accretion results in an increase in the drag force for subsonic perturbers while a diminishment in the force for slightly supersonic perturbers. 
    \item \textbf{Other Forces:} Any perturber with a physical surface will experience a hydrodynamical drag \citep[see Eq.~6 of][]{Villaver_2009}, while massive perturbers with separations on the order of $\lesssim 100's$ of gravitational radii will lose energy and momentum through gravitational wave radiation \citep[\textit{e.g.},][]{Samsing20}. If the perturbers have intrinsic spin, they will experience out-of-plane Magnus and lift forces \citep{DysonLifts}, while the presence of a tertiary companion will alter the orbital dynamics \citep{DodiciTremaine, GaiaMergersInAGN, 2024A&A...683A.135T}. 
\end{itemize}

\subsection{Potential Applications}
The findings presented in this study provide insights into the gas dynamics and orbital evolution of massive perturbers embedded in gaseous media. In this section, we discuss a number of potential astrophysical applications for these results.\\

Binary formation processes are extensively studied with direct applications ranging across many astrophysical systems; ranging from stellar binaries in molecular clouds \citep{Tobin_2018, Cournoyer-Cloutier24}, planetesimals in protoplanetary disks \citep{Nesvorný_2021}, objects in the Kuiper belt \citep{goldreich2002formation}, black hole binary formation in stellar clusters \citep{PortegiesZwart2000,Mapelli21}, and compact objects in active galactic nuclei \citep{BinaryFormationAGN, GDF_BinaryFormationMechanism, Qian2024, DodiciTremaine, 2024MNRAS.535.3630F, 2025MNRAS.tmp..532R}.\\

For applications to perturbers embedded in astrophysical disks, e.g. protoplanetary or active galactic nuclei, the results obtained in this work must satisfy Eq.~(\ref{ShearCondition}) to justify the Green's function solution for a background Keplerian flow. This condition may be achieved for impact parameters $b$ far smaller than the scale height of the disk,
\begin{equation}
    \frac{R_\mathrm{max}}{\sqrt{e^2-1}}\frac{b}{H}\ll 1
\end{equation}
We also note that this condition is independent of the actual orbital dynamics, and thus applies to rectilinear, hyperbolic, elliptical or Jacobi captures. Our findings suggest that following gas-assisted binary formation, the pericenter Mach number may be large. Additionally, the timescale required for these binaries to form may be longer than predicted by the rectilinear proxy model \citep{O99}, as hyperbolic encounters typically lose angular momentum at a rate slower than otherwise expected due to their interactions with Mach cones (see Fig.~\ref{fig:VectorDifference}).\\

Unlike bound systems, hyperbolic scatterings are not cyclic in nature, negating the ability to trace orbital evolution tracks in Fig.~\ref{fig:FlowPlot} \citep[cf. Figure 12 of][where elliptical orbits trace paths over secular timescales]{Paper1}. An alternative approach is to consider the statistical ensemble of equal-mass encounters embedded within a gas-rich cluster \citep{Rozner24b}. Weak-scatterings ($e\gg1$) dominate the cluster dynamics \citep[cf.~Section 7.4.2 of][]{GalacticDynamics} which could be modeled using a deterministic drift coefficient to describe the phase-space evolution of cluster states. This evolution, captured by the distribution function $f(\mathbf{x}, \mathbf{v}, t)$, provides a comprehensive description of the gas-rich cluster over time. Understanding the evolution of nuclear clusters holds implications for supermassive black hole growth as nuclear star clusters enable infall of stars and gas \citep{Stone17, Konstantinos24}, giant molecular clouds \citep{Howard18} and episodes of gas-enriched globular clusters \citep{GDF_inGlobularClusters} as environments for gravitational-wave mergers.\\

Finally, the semi-analytical results presented in this paper should be recoverable through numerical hydrodynamical simulations. In the limit of low-mass perturbers, the density wake morphology and force profiles should match the findings of this paper, enabling future studies to explore the non-linear effects of hyperbolic encounters in gaseous media.

\appendix

\section{Applications in Astrophysical Disks}
\label{ApplicationsInAstrophysicalDisks}

In this paper, we assumed that the background medium is both homogeneous and static as a simplifying assumption. In this appendix, we present a brief discussion on the applicability of our results to astrophysical accretion disks. In such a case, differential rotation, the Coriolis force and the tidal force may have an impact on the density wake structure, the dynamical forces and the orbital evolution results found in this work. To quantify the regimes in which it is appropriate to employ our findings, we investigate the conditions under which it is reasonable to approximate the background medium as being both homogeneous and static. We consider a local approximation to write the continuity and momentum equations for the local shearing sheet as
\begin{align}
    \frac{\partial\rho}{\partial t} + \nabla_j(\rho{v}^j) &= 0,\label{Continuity}\\
    \rho\frac{\partial v^i}{\partial t} + \rho v^j\nabla_jv^i =-\nabla^iP-2\rho\Omega_0 &(\hat{e}_z\times \vec{v})^i-\rho\nabla^i\Phi_\mathrm{T}-\rho\nabla\Phi_\mathrm{pert},\label{Momentum}
\end{align}
where $\Phi_\mathrm{T}=q_\mathrm{s}\Omega_0^2x^2$ is the tidal potential, $x$ is the `Cartesian' shearing sheet coordinate, $q_\mathrm{s}$ is the shear parameter and $\Omega_0$ is the angular frequency of the corotating frame. We expand Eqs.~(\ref{Continuity}) and (\ref{Momentum}) to first order in the density and velocity perturbations, $\rho=\rho_0(1+\alpha), v^i = v_0^i + \beta^i c_\mathrm{s}$, where $v_0^i$ is the background velocity of the gas as measured from the corotating frame. The linearised continuity and momentum equations in the shearing sheet may be written as,
\begin{align}
    \frac{\partial\alpha}{\partial t} + c_\mathrm{s}&\nabla_j{\beta}^j = 0,\label{LinearContinuity}\\
    c_\mathrm{s}\frac{\partial \beta^i}{\partial t} +c_\mathrm{s}^2\nabla^i\alpha+v_0^jc_\mathrm{s}\nabla_j\beta^i+\beta^jc_\mathrm{s}&\nabla_jv_0^i+2\Omega_0c_\mathrm{s}\rho_0(\hat{e}_z\times \boldsymbol{\beta})^i=-\nabla\Phi_\mathrm{pert}.\label{LinearShearingMomentum}
\end{align}
By taking the time derivative of Eq.~(\ref{LinearContinuity}) and the divergence of Eq.~(\ref{LinearShearingMomentum}), we can combine both into a single equation,
\begin{equation}
    -\frac{1}{c_\mathrm{s}^2}\frac{\partial^2\alpha}{\partial t^2}+\nabla^2\alpha+\frac{1}{c_\mathrm{s}}\left[v_0^j\nabla_j\nabla_i\beta^i+2\nabla_iv_0^j\nabla_j\beta^i\right] - \frac{2\Omega_0\hat{e}_z\cdot(\nabla\times\boldsymbol{\beta})}{c_\mathrm{s}} = -\frac{1}{c_\mathrm{s}^2}\nabla^2\Phi_\mathrm{pert}.
    \label{CombinedEquation}
\end{equation}
We recognise the first two terms in Eq.~(\ref{CombinedEquation}) as the Green's function propagator, whose solutions are described in this paper. In the limit where all other terms on the left hand-side are negligible, it is accurate to employ the solutions obtained in this work.\\

We first assume that the perturbations in density and velocity are of the same order, $\beta^i\sim\alpha$. Secondly, based on Eq.~(\ref{GeneralSolutionAlpha}) we employ a simplified scaling relation for $\alpha$ (and thus $\beta^i$) given by \begin{equation}
    \alpha(x), \beta^i(x) \sim  \frac{r_\mathrm{B}}{|x|},
\end{equation}
for which the derivatives of perturbed quantities thus scale as $\nabla^n\alpha \sim r_\mathrm{B}/|x|^{n+1}$ where $r_\mathrm{B}\equiv Gm/c_\mathrm{s}^2$ is the Bondi radius. Thirdly, we now only consider the free propagation of sound waves far away from the perturber, for which we drop the source term $\nabla^2\Phi_\mathrm{pert}$. Finally, we seek to bring Eq.~(\ref{CombinedEquation}) into dimensionless form by multiplying across by $r_\mathrm{B}^2$, giving
\begin{align}
    -\frac{r_\mathrm{B}^2}{c_\mathrm{s}^2}\frac{\partial^2\alpha}{\partial t^2}+r_\mathrm{B}^2\nabla^2\alpha = 
    -\frac{q_\mathrm{s}\Omega_0 x}{c_\mathrm{s}}\left[\left(\frac{v_0^j}{q_\mathrm{s}\Omega_0x}\right)r_\mathrm{B}^2\nabla_j\nabla_i\beta^i
    +\left(\frac{2\nabla_iv_0^j}{q_\mathrm{s}\Omega_0}\right)\frac{r_\mathrm{B}^2\nabla_j\beta^i}{x}
    \right]
    +\frac{2\Omega_0x}{c_\mathrm{s}}\left[\frac{r_\mathrm{B}^2}{x}\hat{e}_z\cdot(\vec{\nabla}\times\boldsymbol{\beta})\right] .
    \label{DimensionlessMasterEq}
\end{align}
We assume that the background velocity is of the order $v_0^j\sim q_\mathrm{s}\Omega_0 x$ and its derivatives $\nabla_j v_0^i\sim q_\mathrm{s}\Omega_0$. On the right-hand side of Eq.~(\ref{DimensionlessMasterEq}) we have constructed the square brackets to be on the order of $\alpha^3$, although weighted according to relevant scales of the system. Therefore, in the limit of
\begin{align}
    \left|\frac{q_\mathrm{s}\Omega_0x}{c_\mathrm{s}}\right|&\ll 1,\label{Cond1}\\
    \left|\frac{\Omega_0x}{c_\mathrm{s}}\right|&\ll 1. \label{Cond2}
\end{align}
the Green's function solution is applicable. Eq.~(\ref{Cond1}) describes the ratio of the advection speed to the sound speed. Reassuringly, this must be small to justify neglecting the effects of shear and thus recovering the Green's function solution for a static, homogeneous medium. Eq.~(\ref{Cond2}) requires that the Coriolis force (final term in Eq.~\ref{DimensionlessMasterEq}) is small. For a rigidly rotating disk ($q_\mathrm{s}=0$) this is the only requirement needed to be satisfied, as might be expected.\\

We require the perturbation wake (of spatial scale $L$) to be contained within the shearing sheet, $x\lesssim L$ (where $L\gg r_\mathrm{B}$). For a Keplerian disk, the shear parameter is of order unity ($|q_\mathrm{s}|=3/2$), for which Eqs.~(\ref{Cond1}) and (\ref{Cond2}) become equivalent conditions. Assuming the disk is in vertical hydrostatic equilibrium with the scale height $H = c_\mathrm{s}/\Omega_0$, we can write the condition for neglecting the effects of shear and the Coriolis force for a Keplerian disk, 
\begin{align}
    \frac{L}{H}&\ll 1.\label{ShearCondition}
\end{align}
We note that for the special case of a density wake which spans the Hill radius $R_\mathrm{H}$ of the perturber ($L = R_\mathrm{H}$), Eq.~(\ref{ShearCondition}) corresponds to the opposite extreme of the ``strong-shock'' limit for gap-opening perturbers \citep[cf.][]{Crida06,Muto10,Duffell13},
\begin{equation}
    \frac{1}{h}\left(\frac{q_*}{3}\right)^{1/3}\gg 1
\end{equation}
where $h = H/R$ is the aspect ratio in the disk and $q_* = M_*/m$ is the mass ratio between the central body and the perturber. When this quantity is large, the perturber will carve a gap in the disk, whereas when it is small the perturber produces density wakes which are described accurately by the Green's function solution as described in this study. Finally, when $L \ll R_H$ the tidal forces from the central body can also be neglected. \\

\section{The ``Two-wake'' Proxy Based on \cite{O99}}
\label{sec:AdvancedProxy}

The ``one-wake'' rectilinear proxy estimates the force exerted on a single perturber by its own wake as it moves along a prescribed hyperbolic orbit, as outlined in Section~\ref{sec:RectlinearProxy}. However, in equal-mass encounters, the wake produced by the companion perturber also influences the total force on each body. In this appendix, we explain how to construct the ``two-wake'' proxy introduced in Section~\ref{sec:RectlinearProxy} and compare its results with those presented in Section \ref{sec:OrbitalEvolution}.\\

\begin{figure}
    \centering
    \includegraphics[width=\linewidth]{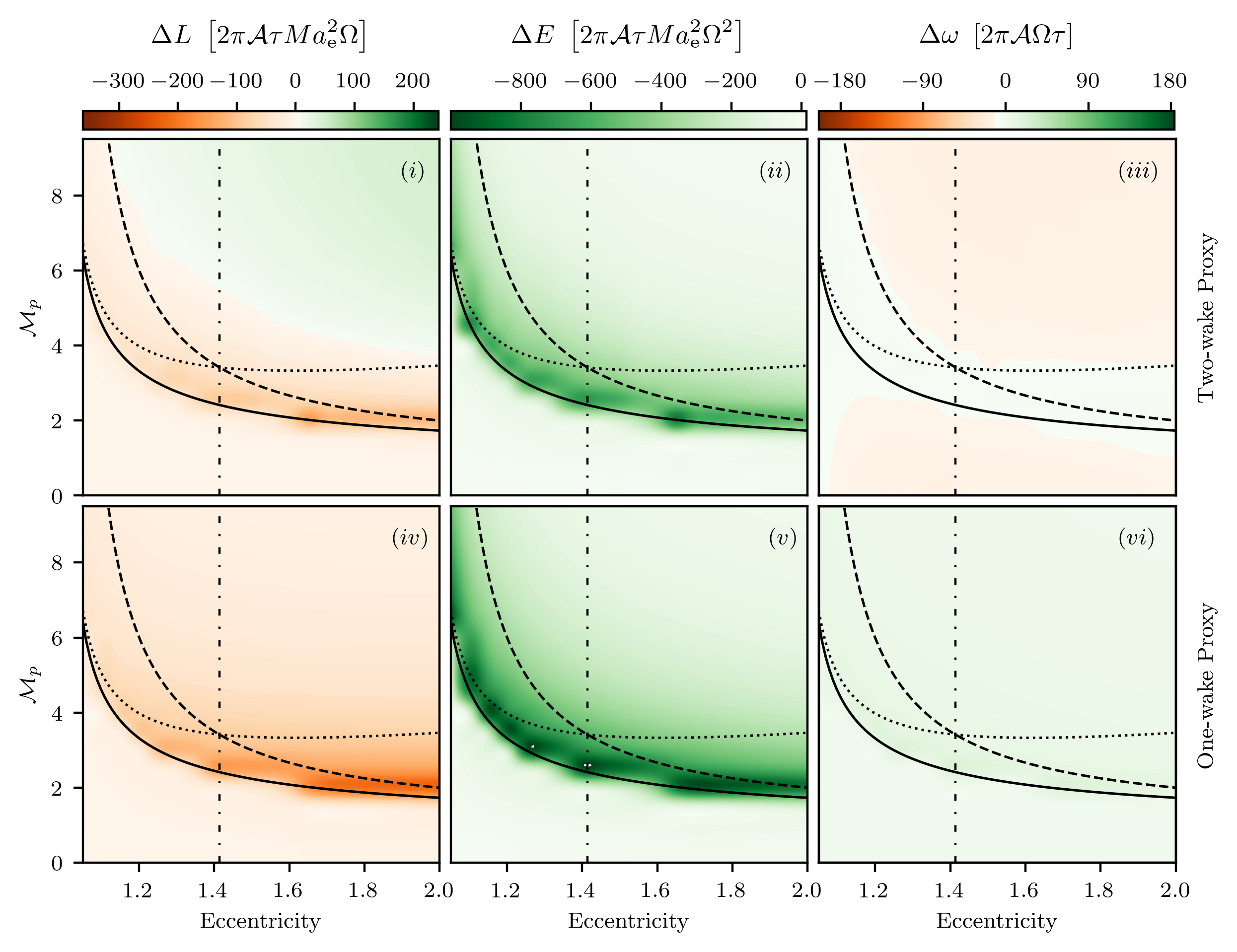}
    \caption{The changes to specific energy, specific angular momentum, and angle of pericenter as a function of the orbital eccentricity and pericenter Mach number. \emph{Top:} Results obtained with the ``two-wake'' proxy model. \emph{Bottom:} Results obtained with the ``one-wake'' proxy model. The solid, dashed and dotted lines indicate the transonic, companion extraction, and self-extraction separatrices, respectively.}
    \label{fig:AdvancedProxy}
\end{figure}

Following the same approach described in Section~\ref{sec:RectlinearProxy}, we decompose the hyperbolic encounter into a series of constant, rectilinear segments. At a given time, each perturber travels with Mach number $\mathcal{M}=V/c$ (see Eq.~\ref{HyperbolicVelocity}) with respect to the medium. We then proceed as follows. First, we consider an axis $\xi$ parallel with the velocity of the perturbers, $\eta$ an axis in the orbital plane perpendicular to the velocity, and $z$ the out-of-plane axis as defined before. In these units, we use Eq.~10 of \citetalias{O99} to write the density wake centered on one perturber (where the perturber's position corresponds to $\xi = \eta = z =0$) as,
\begin{align}
    \alpha_\mathrm{O99}(t, \eta, \xi, z) = \frac{GM}{c_\mathrm{s}^2}\frac{1}{\sqrt{(\xi-Vt)^2 + R^2(1-\mathcal{M}^2)}}\begin{cases}
        1 & \mathrm{if}~~R^2+\xi^2<(c_\mathrm{s}t)^2\\
        2 & \mathrm{if}~~\mathcal{M}>1, R^2+\xi^2>(c_\mathrm{s}t)^2, (\xi-Vt)/R < -\sqrt{\mathcal{M}^2-1}, \xi>c_\mathrm{s}t/\mathcal{M}\\
        0 & \mathrm{otherwise}
    \end{cases}
\end{align}
where $R^2 = (\eta^2+z^2)$. Second, we consider the relative position of the companion perturber in Cartesian coordinates as $\Delta\mathbf{x}_{r} = (\Delta x, \Delta y)$, with
\begin{align}
    \Delta x &= -2\cosh\sigma(t) - 2e,\\
    \Delta y &= -2\sqrt{e^2-1}\sinh\sigma(t).
\end{align}
Third, by performing a rotation of angle $T$ in the $xy$-plane (see Fig.~\ref{fig:Schematic}), we obtain the tangential and perpendicular components of the relative position vector, providing the location of the companion at position $(\Delta \eta, \Delta \xi)$. Finally, we note that the entire structure of the companion wake is flipped in orientation so we can obtain the density structure to the ``two-wake'' proxy as,  
\begin{equation}
    \alpha_\mathrm{two-wake} = \frac{1}{2}\alpha_\mathrm{O99}(t, \eta, \xi, z) + \frac{1}{2}\alpha_\mathrm{O99}(t, \Delta\eta -\eta, \Delta\xi -\xi, z),
\label{AdvancedProxyWake}
\end{equation}
In order to obtain the changes to the orbital energy, angular momentum, and argument of pericenter for a given perturber, we numerically integrate the ``two-wake'' density structure given in Eq.~(\ref{AdvancedProxyWake}), to obtain the ``two-wake'' force exerted on it, followed by the coordinate transformation in Eq.~(\ref{TangentialPolar}). We present our results for ``two-wake'' proxy on the top row of Fig.~\ref{fig:AdvancedProxy}, with the ``one-wake''  proxy on the bottom row.\\

\acknowledgements
We thank Scott Tremaine, Philip Kirkeberg, Chris Hamilton and Johan Samsing for useful discussions. D.ON and D.J.D. acknowledge support from the Danish Independent Research Fund through Sapere Aude Starting Grant No. 121587. 
D.J.D. received funding from the European Union's Horizon 2020 research and innovation programme under Marie Sklodowska-Curie grant agreement No. 101029157
M.E.P. gratefully acknowledges support from the Independent Research Fund Denmark via grant ID 10.46540/3103-00205B and the Institute for Advanced Study, where part of this work was carried out. 
This work was supported in part by ERC Starting Grant no. 101043143.
The Tycho supercomputer hosted at the SCIENCE HPC center at the University of Copenhagen was used to support this work.

\bibliography{bibliography.bib}

\end{document}